\documentclass[12pt]{article}
\usepackage{newtxtext,newtxmath}
\usepackage{graphicx}
\usepackage{multirow}
\usepackage{amsmath}
\usepackage{mathrsfs}
\usepackage{textcomp}
\usepackage{booktabs,longtable,array}
\usepackage{calc}
\usepackage{adjustbox}
\usepackage[normalem]{ulem}
\usepackage[letterpaper,margin=1in]{geometry}
\usepackage[numbers,sort&compress,super]{natbib}
\usepackage{url}
\usepackage[colorlinks=true,linkcolor=blue,citecolor=blue,urlcolor=blue]{hyperref}
\usepackage[capitalise,nameinlink]{cleveref}
\usepackage{totcount}
\usepackage{bibunits}
\defaultbibliographystyle{sciencemag}
\defaultbibliography{references}
\graphicspath{ {./figures/} }
\regtotcounter{figure}
\regtotcounter{table}
\newtotcounter{suppnote}
\date{}
\renewenvironment{abstract}{\quotation}{\endquotation}
\renewcommand\refname{References}
\makeatletter
\renewcommand{\fnum@figure}{\textbf{Figure \thefigure}}
\renewcommand{\fnum@table}{\textbf{Table \thetable}}
\makeatother
\newcommand{\backmatter}{}
\newcommand{\figurecaptionpage}[2]{%
  \clearpage
  \refstepcounter{figure}%
  \noindent\textbf{Figure~\thefigure:} #1\label{#2}\par
}
\newcommand{\suppnote}[1]{%
  \stepcounter{suppnote}%
  \section*{Supplementary Note \thesuppnote. #1}%
}
\def\scititle{Scientific exploration, collaboration and labor division in the large language model era}
\title{\bfseries \boldmath \scititle}
\author{Xiang~Zheng$^{1}$, Xi~Hong$^{1}$, Jialin~Liu$^{1}$, Chaoqun~Ni$^{1\ast}$\and
  \small$^{1}$Information School, University of Wisconsin--Madison, Madison, WI, USA.\and
  \small$^\ast$Corresponding author. Email: chaoqun.ni@wisc.edu}

\begin{document}
\begin{bibunit}[sciencemag]
  \maketitle
  \begin{abstract} \bfseries \boldmath
    Large language models (LLMs) have rapidly and significantly entered scientific workflows, but it remains unclear how their diffusion is associated with changes in scientists' strategies in research directions and team building. We link PubMed Central full text with OpenAlex publication and collaboration histories for 775,323 scientists and analyze CRediT contribution statements from 137,120 multi-author papers. After 2022, scientists increasingly published across more intellectually distant fields and entered fields in which they had not previously worked. These increases in interdisciplinarity and exploration were especially pronounced among established scientists and scientists from non-English-speaking low- and middle-income countries.
    Authors with stronger AI-writing signals were already more interdisciplinary and exploratory before the widespread adoption of LLMs, and the gap widened more after 2022 compared to authors with less AI-writing signals. Scientists' collaboration networks also became more interdisciplinary after 2022. Yet, among authors with stronger AI-writing signals, research interdisciplinarity was less closely tied to the disciplinary diversity of their collaborators.
    The division of labor within research teams also became more differentiated. Contributors on papers published after 2022 reported narrower role sets on average, coauthors shared fewer roles in common, and their role profiles became less rigid and more fluid. Roles such as software and validation tasks increased, while conceptual and managing roles decreased. These patterns suggest that team members are taking on more distinct responsibilities and may rely less on one another to perform research tasks.
    Overall, this study reflects that the LLM era coincides with a broader reorganization of scientific exploration, collaboration, and the division of labor.
  \end{abstract}


  \section{Introduction}
  \label{sec:introduction}

  The rise of large language models (LLMs) and related artificial intelligence
  (AI) tools has profoundly influenced how knowledge is created and shared
  \cite{RiseLargeLanguage2025}. Since the public diffusion and large-scale
  application of LLM after the release of ChatGPT in November 2022 and GPT-4 in
  March 2023, LLM tools have been rapidly evolving and exhibit great advantages
  in tasks such as writing, coding, brainstorming, and literature search
  \citep{Doshi2024Generativea,Lee2024empirical}. Some studies suggest LLMs even
  have the potential to enhance human creativity and generate novel ideas
  \cite{ICLR2025_ea94957d, yangCanLargeLanguage2026}. Scientists quickly began
  using LLMs and AI in grant work and manuscript preparation
  \citep{Stokel2023What}, and the growing use of LLMs in writing appears to be
  difficult to curb through restrictive measures such as journal policies
  \citep{heAcademicJournalsAI2026}. Surveys and empirical studies show that LLM
  use in research has had a sharp increase since 2023 and is widespread across
  tasks but is uneven across fields, language contexts, institutions, and career
  stages
  \citep{chugunovaWhoUsesAI2026,KobakDelving,liangQuantifyingLargeLanguage2025,liuAIAssistedWritingGrowing2025,silerDiffusionLargeLanguage2026,VanNoorden2023AI,
    Gao2024Quantifying}. Other work links LLM to productivity gains, faster
  scientific production and reduced language barriers
  \citep{kusumegiScientificProductionEra2025,Noy2023Experimental}.

  Meanwhile, beyond the rapid increase in scientific production, it remains
  unclear whether and how scientists' strategies in research directions and team
  building changed with LLM diffusion. When setting research directions,
  scientists constantly face the trade-off of exploitation and exploration.
  Exploitation refines known lines of work, while exploration may open
  possibilities of alternative choices and new problem spaces
  \citep{March1991Exploration}. Through exploration of new research agendas,
  scientists may reorganize the direction of their work and engage with new
  areas, which may produce high-impact insights
  \citep{Hill2025pivota,Tripodi2025Tenureb}. Recent advances in AI for science
  may further expand such opportunities by providing new tools for hypothesis
  generation, simulation, prediction, data analysis, and experimental design,
  thereby making previously inaccessible questions or methods more feasible and
  enabling scientists to explore or pivot toward new research directions.
  However, exploration and interdisciplinary research can also be risky, as
  scientists who move into unfamiliar areas must learn new vocabularies, assess
  unfamiliar standards of evidence, find audiences, and negotiate review
  expectations, which increase the burden of knowledge and learning costs
  \citep{Jones2009Burden, Hill2025pivota, Leahey2017Prominent}. LLMs may reduce
  some of these barriers by helping scientists navigate unfamiliar literature,
  terminology, code, and methods, and more AI-enabled scientific tools may
  directly broaden the range of questions they can investigate
  \citep{liDecipheringScientificCollaboration2025,xuAdaptingLLMsHow2025}. Whether
  these capabilities translate into greater exploration, interdisciplinary
  movement, and research pivoting in practice, however, remains an empirical
  question.

  Exploratory and boundary-crossing work is also social. To access expertise,
  data practices, instruments, and validation norms outside their own fields,
  scientists may build teams that distribute expertise across collaborators
  \citep{Hall2018science,Lariviere2015Longdistance, Salatino2026Does}.
  Nevertheless, cross-disciplinary collaboration has been challenging because of
  the barriers between disciplinary norms and cultures
  \citep{Lariviere2015Longdistance}, which LLMs can be expected to translate and
  improve the communication across multiple disciplines. Moreover, more technical
  techniques, especially LLM- and AI-related methods, have entered various fields
  of science, and such work often require interdisciplinary collaboration with
  computational, data, and validation expertise
  \cite{liAcademicCollaborationLarge2024}. LLM diffusion may therefore be
  associated not only with what scientists study, but also with whom they work.

  Team organization also depends on how collaborators share labor. Cross-field
  projects often require collaborators with different skills and expertise to
  contribute together; these ``team-players'' are shown to be the majority within
  team collaboration \citep{Lu2020Cocontributorship}. On the other hand,
  LLM-driven automation tools can let scientists take on tasks that previously
  required more direct human assistance \citep{Boiko2023Autonomous}. LLMs and AI
  agentic systems exhibit the potential to lower the cost of failure, making
  riskier and more ambitious ideas more practical for individuals to test when
  they were once too costly or time-consuming \citep{wangAIAgentsAre2026}. LLMs
  may also lower the cost of integrating and coordinating specialized tasks
  performed by individuals within a team
  \citep{Burton2024How,Woolley2026Generative}. These changes make reported
  contribution roles another place to look for shifts in the organization of
  scientific work, even when team membership itself does not fully capture such
  shifts.

  Nevertheless, less large-scale empirical evidence documents whether the LLM
  diffusion is associated with these linked organizational changes: broader
  scientific exploration in authors' field portfolios, more diverse collaborator
  pools, and changes in labor division inside teams. Related studies examined
  papers applying GenAI and LLM-related techniques, finding shifts in research
  focuses, distinctive collaboration, international participation, and
  disciplinary patterns
  \citep{Ding2025Rise,liDecipheringScientificCollaboration2025,xuAdaptingLLMsHow2025,
    haoArtificialIntelligenceTools2026}. Recent studies have also implied ongoing
  changes in team sizes and authors' country compositions after 2022
  \cite{Ben-Zion2026Changes, Yan2026AIassisted}. These studies are important but
  do not fully answer whether broad scientific populations, not just LLM
  researchers but also LLM adopters, changed how they move across fields, with
  whom they collaborate, and how they allocate tasks after LLMs became widely
  available.

  To fill in the gap, we examine a multilevel reorganization of scientific work
  in the LLM era: greater breadth in what scientists study, greater disciplinary
  diversity in whom they work with, and greater differentiation in how team
  members report their contributions. Empirically, we combine full-text PubMed
  Central articles with OpenAlex author histories to reconstruct author
  portfolios and collaboration profiles within primarily broad biomedical fields.
  We then combine PLOS and PMC articles with structured CRediT contribution
  statements to examine reported roles within papers. The design is descriptive
  and is not meant to estimate the causal effect of LLM adoption, but the purpose
  is to show whether the period after broad public LLM diffusion, and the papers
  and authors with stronger LLM-like writing signals, are associated with
  systematic changes in exploration, collaboration, and reported labor division.

  \section{Results}
  \label{sec:results}

  \subsection{Research portfolios and exploration expanded after 2022}

\begin{figure}[htbp]
    \centering
    \includegraphics[width=1\textwidth]{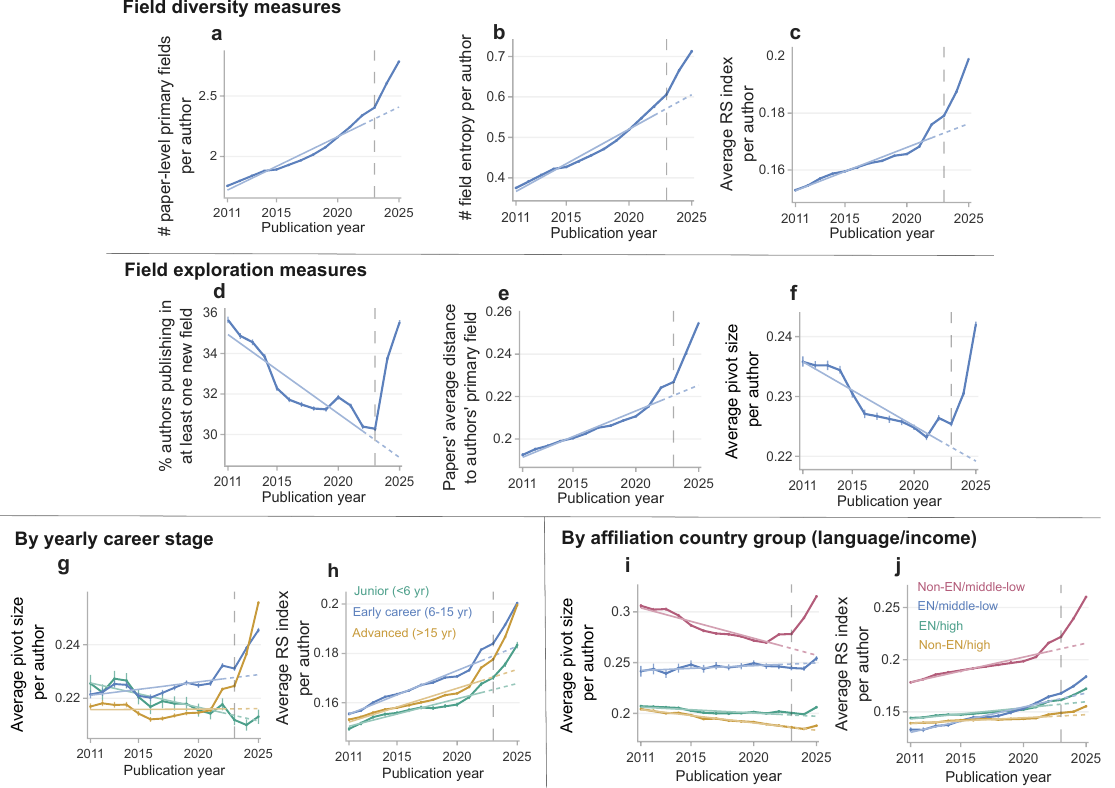}
    \caption{\textbf{Annual expansion of research portfolios and heterogeneity.} Dashed vertical lines mark 2023. Solid trend lines are linear fits estimated from 2011--2022 values; dashed extensions indicate expected post-2022 values under pre-2023 trends. Error bars denote 95\% bootstrap confidence intervals. (a) Average number of primary fields in which authors publish. (b) Field-based Shannon entropy of authors' publication portfolios. (c) Rao--Stirling interdisciplinarity index based on primary fields. (d) Share of authors publishing at least one paper in a field in which they have not previously published. (e) Average field-level distance between papers and authors' primary fields. (f) Average pivot size from authors' recent reference profiles. (g) Average pivot size by yearly career stage. Years in brackets denote the publication history range up to the focal year. For junior scholars, we only include those with at least 3-year publication history for stable pivot calculation. (h) Average Rao--Stirling index by yearly career stage. (i) Average pivot size by affiliation-country language and income group. EN: English-speaking countries. (j) Average Rao--Stirling index by affiliation-country language and income group.}
    \label{fig:portfolio-expansion}
  \end{figure}

  After 2022, scientists increasingly published across a broader range of fields.
  The average number of distinct primary fields represented in scientists' papers
  increased from 2.3 in 2022 to 2.8 in 2025, reaching a level 15.6\% higher than
  would be expected based on the continuation of the pre-2023 trend
  (\cref{fig:portfolio-expansion}a). To account for the distribution of papers
  across primary fields, we calculated Shannon entropy, a widely used diversity
  index, which increased from 0.55 in 2022 to 0.71 in 2025, showing that
  scientists' research is less concentrated in a single field
  (\cref{fig:portfolio-expansion}b). The Herfindahl-Hirschman Index (HHI),
  another measure of diversity, shows a similar pattern (Supplementary
  \cref{si-fig:si-alternative-measures}b). We further incorporated knowledge
  proximity between fields by calculating the Rao-Stirling index, a widely used
  measure of interdisciplinarity, based on papers' primary fields (Supplementary
  \cref{si-fig:si-alternative-measures}c for results based on reference fields).
  This measure also increased substantially after 2022, rising from 0.176 in 2022
  to 0.199 in 2025; by 2025, it was 13.0\% higher than the value expected under
  the pre-2023 trend. References also became more field-spanning: the share of
  references outside the citing paper's field rose to 37.4\% in 2025, 6.1\%
  higher than the expected value (Supplementary
  \cref{si-fig:si-alternative-measures}a).

  Scientists also intensified their exploration of new fields and research
  directions after 2022. In 2025, 35.5\% of authors published in at least one
  field in which they had not previously published, compared with 30.4\% in 2022,
  and this level was 23.2\% higher than would be expected based on the pre-2023
  trend (\cref{fig:portfolio-expansion}d). The average share of papers published
  outside an author\textquotesingle s primary field likewise increased from
  39.7\% in 2022 to 44.6\% in 2025 (Supplementary
  \cref{si-fig:si-alternative-measures}e). Consistent with this pattern, the
  average distance between a paper's field and the author's primary field rose to
  0.255 in 2025, 13.0\% above the value expected from the pre-2023 trend
  (\cref{fig:portfolio-expansion}e). Scientists' top field share also followed a
  consistent pattern (Supplementary \cref{si-fig:si-alternative-measures}d). The
  number of new primary fields an author enters for the first time showed the
  same post-2022 increase (Supplementary \cref{si-fig:si-alternative-measures}f).

  We further examined whether scientists were venturing beyond their established
  areas of expertise by measuring the extent to which a paper's references
  pivoted from the author's recent referencing profile. This pivot measure
  exhibited a rebound pattern, declining from 0.236 in 2011 to 0.226 in 2022
  before increasing to 0.242 in 2025 (\cref{fig:portfolio-expansion}f). In
  addition, by constructing individual-level co-reference networks
  \cite{Zeng2019Increasing}, we identified the evolution of scientists' research
  agendas over time and found that a growing share of papers published after 2022
  marked the emergence of new research agendas (Supplementary
  \cref{si-fig:si-new-agenda}). We confirmed that the above patterns are robust
  to alternative disciplinary classifications (Supplementary
  \cref{si-fig:si-field-definitions}). The same patterns remain when
  low-productivity scientists are excluded (Supplementary
  \cref{si-fig:si-restricted-sample}). The post-2022 increase is still evident
  after controlling for publication counts and fixed effects (Supplementary
  \cref{si-fig:si-event-study}).

  This shift diversified the topics and cross-field research after 2022: The Gini
  index of the paper distribution across over 3,000 OpenAlex topics declined in
  major fields after 2022, indicating that publications became less concentrated
  in a small set of topics and more widely distributed across research topics
  (Supplementary \cref{si-fig:si-topic-diversity}a). The coefficient of variation
  in the destinations of authors' out-of-field publications fell after 2022,
  showing that authors from the same primary field were spreading their work more
  evenly across a wider set of other fields rather than moving toward only a few
  common destinations (Supplementary \cref{si-fig:si-topic-diversity}b).
  Field-pair comparisons between 2020--2022 and 2023--2025 further show that more
  scientists published concurrently in biomedical fields and technical fields,
  especially engineering, while fewer authors remained within adjacent biomedical
  field combinations (Supplementary \cref{si-fig:si-topic-diversity}c). At the
  subfield level, more authors published in Biomedical Engineering, Electrical
  and Electronic Engineering, and Mechanical Engineering concurrently with
  biomedical fields after 2022, while non-Engineering subfields, such as
  Materials Chemistry, Molecular Biology, Artificial Intelligence, and Renewable
  Energy also showed substantial increases (Supplementary
  \cref{si-fig:si-topic-diversity}d).

  The portfolio changes are not uniform across scientists. By career stage, the
  clearest increase in pivot size appears among advanced authors (from 0.223 in
  2022 to 0.256 in 2025). In contrast, junior authors' pivot remained relatively
  stable (\cref{fig:portfolio-expansion}g). The career-stage pattern is also
  visible for the Rao--Stirling index (\cref{fig:portfolio-expansion}h). Country
  and income groups show a second heterogeneity pattern. Non-English low- and
  middle-income-country authors have the highest level and the clearest post-2022
  increase in pivot, rising from 0.270 to 0.315
  (\cref{fig:portfolio-expansion}i). Their average Rao-Stirling index also rises
  from 0.216 to 0.260 (\cref{fig:portfolio-expansion}j). In contrast, English
  high-income-country authors and English low/middle-income-country authors show
  rising Rao-Stirling but relatively flat pivot. This is most likely driven by
  Chinese authors, as the rise in the two selected measures is most significant
  compared with other major countries in our sample (Supplementary
  \cref{si-fig:si-top-country-trends}). Moreover, disaggregated by authors'
  primary topical domains, while the pattern holds across different domains,
  Physical Sciences (including Engineering, Computer Science, Chemistry,
  Mathematics, Physics, Earth, etc.) are the most prominent (Supplementary
  \cref{si-fig:si-domain-trends}).

  \subsection{AI-writing intensity and portfolio breadth}

\begin{figure}[htbp]
    \centering
    \includegraphics[width=1\textwidth]{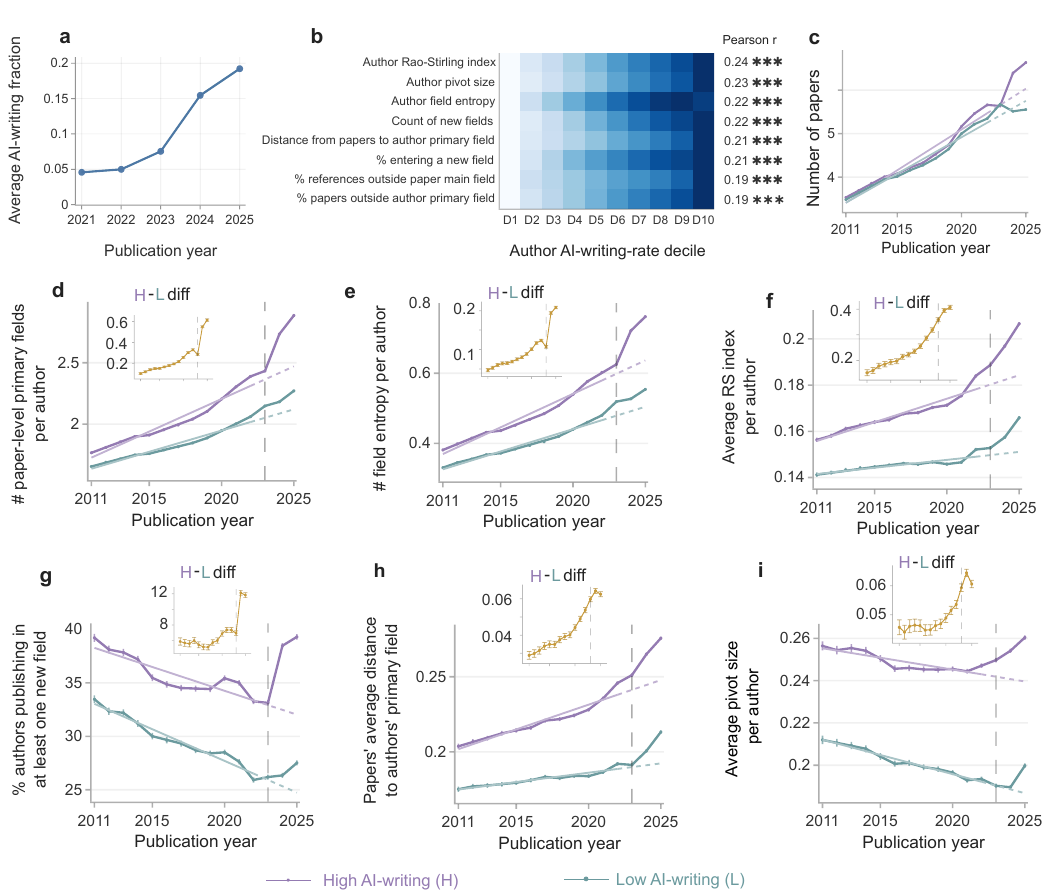}
    \caption{\textbf{Authors' portfolio differences.} Dashed vertical lines mark 2023. Solid trend lines are linear fits estimated from 2011--2022 values; dashed extensions indicate expected post-2022 values under pre-2023 trends. Error bars denote 95\% bootstrap confidence intervals. (a) Average paper-level AI-writing fraction by publication year. (b) Correlation heatmap and correlation coefficients linking author AI-writing-rate deciles with portfolio and exploration measures. *** $p<0.001$. (c)-(i) Field diversity and exploration, same as Figure 1a-f, for high- and low-AI-writing groups. Insets show the high-minus-low differences. Authors are weighted by their author feature strata derived from coarsened exact matching.}
    \label{fig:ai-writing-portfolios}
  \end{figure}

  In addition to the temporal comparisons, we followed prior studies and computed
  an AI-writing fraction measure for each paper according to the extent to which
  their language resembles LLM-assisted writing (Supplementary Note 1). A survey
  on 816 scientists suggests that, compared with scientists who report not using
  AI for writing, those who use AI for writing are 2.2 to 5.6 times more likely
  to also use AI for other research tasks, such as information search, idea
  framing, and analytical support \citep{Liao2024LLMs}. This measure, therefore,
  plausibly reflects but does not directly indicate broader exposure to LLMs in
  the research process. The average AI-writing fraction based on all 2021-2025
  publications in PubMed Central (PMC) increased substantially from 2021 (0.05)
  to 2025 (0.19), indicating the growing popularity of LLM-assisted writing after
  2022 (\cref{fig:ai-writing-portfolios}a). Among the authors, young scientists,
  engineering-related scientists, and those from non-English-speaking countries
  have higher AI-writing rates, which aligns with previous studies' estimates
  \citep{liuAIAssistedWritingGrowing2025,silerDiffusionLargeLanguage2026}.

  We compute a scientist's AI-writing rate at the author-level by averaging their
  papers' available AI-writing fraction values in the post-2022 window.
  Scientists' AI-writing rate is strongly associated with measures of
  interdisciplinarity and research exploration. For example, the Rao-Stirling
  index is positively correlated with the AI-writing rate ($r=0.237, p<0.001$).
  Authors in the first decile of the Rao-Stirling distribution have a mean
  AI-writing rate of 0.143, whereas those in the tenth decile have a mean rate of
  0.230. Similarly, pivot size increases from 0.172 in the first decile to 0.295
  in the tenth decile and is also positively correlated with the AI-writing rate
  ($r=0.230, p<0.001$; \cref{fig:ai-writing-portfolios}b).

  We next compared high- and low-AI-writing authors. We classified authors into
  high- and low-AI-writing groups. Authors were assigned to the high-AI-writing
  group if the average AI-writing fraction of their papers published between 2023
  and 2025 exceeded 0.15 ($n=244,062$); and the low-AI-writing group contains
  those whose average AI-writing fractions are below 0.05, similar to the
  pre-2023 level ($n=211,911$; \cref{fig:ai-writing-portfolios}c). These cutoffs
  are anchored to the 2021--2022 paper-level baseline. The yearly paper-level
  means are 0.0457 and 0.0499; thus, we use 0.05 as the rounded non-LLM-writing
  baseline and the diagnostic expectation for authors without AI writing
  assistance. The 0.15 threshold identifies an author-level rate at least three
  times that baseline, reducing the chance of false positive cases.

  To improve comparability between the two groups, we matched authors annually on
  exact career stage, primary field, affiliation country group, and binned levels
  of productivity over the focal year and the preceding two years. The matched
  comparisons show that high-AI-writing authors became more productive after
  2022, whereas their pre-2023 productivity was similar to that of low-AI-writing
  authors. However, high-AI-writing authors had been more interdisciplinary and
  exploratory: they worked in more fields (\cref{fig:ai-writing-portfolios}d-f)
  and also continued to explore new fields before and after 2023 than the low
  AI-writing group (\cref{fig:ai-writing-portfolios}g-i). The pre-2023
  differences indicate that high-AI-writing authors are not a random subset of
  the scientific population. Rather, they already had broader portfolios before
  the widespread diffusion of LLMs. Nevertheless, the gaps between high- and
  low-AI-writing authors continued to widen after 2023, which still holds after
  controlling author-level fixed effects and publication counts (Supplementary
  \cref{si-fig:si-high-low-event}). Thus, the post-2022 association between
  AI-writing rate and exploration likely reflects a selection-plus-reinforcement
  pattern into LLM-assisted writing and research behavior.

  Although the AI-writing proxy helps identify authors and papers associated with
  wider exploration, it cannot identify a causal channel. High-AI-writing authors
  may be early adopters of general-purpose tools; they may also be more
  productive, better resourced, more interdisciplinary or more internationally
  connected before LLMs entered the workflow. The evidence supports the claim
  that AI-writing intensity and exploration correlate together, not that LLM use
  alone produced the observed exploration.

  \subsection{Collaboration networks became more interdisciplinary}

  The preceding analyses show that scientists increasingly publish across
  disciplinary boundaries and explore new research directions. One potential
  explanation is that scientists are drawing on more diverse collaborators. We
  therefore examine whether the observed expansion of research portfolios is
  accompanied by changes in the disciplinary composition of scientists'
  collaboration networks.

\begin{figure}[htbp]
    \centering
    \includegraphics[width=0.9\textwidth]{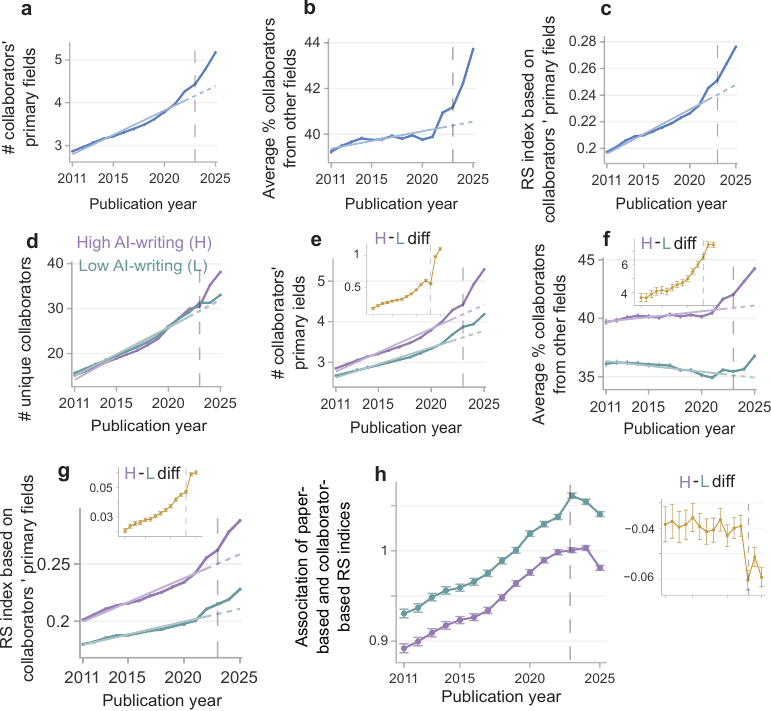}
    \caption{\small \textbf{Collaborators' fields diversified further after 2022.} Dashed vertical lines mark 2023. Solid trend lines are linear fits estimated from 2011--2022 values; dashed extensions indicate expected post-2022 values under pre-2023 trends. Error bars denote 95\% confidence intervals. (a) Average number of distinct primary fields represented among collaborators by year. (b) Average share of collaborators from fields different from the focal author's primary field. (c) Rao-Stirling index based on collaborators' primary fields. (d) Average number of unique collaborators for high- and low-AI-writing author groups. (e) Average number of distinct collaborators' primary fields for high- and low-AI-writing groups, with the inset showing the high-minus-low difference. (f) Average share of collaborators from other fields different from the focal author for high- and low-AI-writing groups, with the inset showing the high-minus-low difference. (g) Collaborator-based Rao-Stirling index for high- and low-AI-writing groups. (h) Association between paper-based and collaborator-based Rao-Stirling indices for high- and low-AI-writing groups. Authors are weighted by author-feature strata derived from coarsened exact matching.}
    \label{fig:collaboration-fields}
  \end{figure}

  Across the sample, scientists increasingly collaborated with scientists from a
  broader range of disciplinary backgrounds after 2022. The average number of
  distinct fields represented among an author\textquotesingle s collaborators
  rose to 5.2 in 2025, compared with an expected value of 4.4 based on the
  continuation of the pre-2023 trend (\cref{fig:collaboration-fields}a). The
  share of collaborators outside an author\textquotesingle s primary field
  increased from 40.9\% in 2022 to 43.7\% in 2025
  (\cref{fig:collaboration-fields}b). Based on collaborators' primary research
  fields, we calculated a collaborator-based Rao--Stirling index, which reached
  0.276 in 2025, approximately 12\% higher than expected under the pre-2023 trend
  (\cref{fig:collaboration-fields}c). At the paper level, the share of papers
  involving at least one cross-field collaborator increased from 70.1\% in 2022
  to 76.0\% in 2025. Together, these results indicate that scientists
  increasingly rely on more diverse disciplinary collaboration networks.

  We next examine whether these patterns differ by AI-writing intensity. Using
  the matched samples described above, we compare high- and low-AI-writing
  authors. High-AI-writing authors had similar numbers of collaborators before
  2023 and only moderately more collaborators afterward
  (\cref{fig:collaboration-fields}d). In contrast, differences in
  interdisciplinary collaboration were already evident before 2023 and widened
  further after that year. High-AI-writing authors consistently collaborated with
  scientists from more distinct fields (\cref{fig:collaboration-fields}e), had a
  larger share of collaborators outside their own primary field
  (\cref{fig:collaboration-fields}f), and exhibited higher collaborator-based
  interdisciplinarity (\cref{fig:collaboration-fields}g). This pattern closely
  parallels the earlier results on publication-based interdisciplinarity and
  exploration. As before, the existence of substantial pre-2023 differences
  suggests that authors with high AI-writing rates were already more
  interdisciplinary before the widespread adoption of LLMs. Consequently, the
  observed post-2022 divergence likely reflects both selection into AI-assisted
  writing and potential changes associated with these tools.

  We further ask how much of an author\textquotesingle s interdisciplinarity can
  be explained by collaborators' interdisciplinarity. We used event-study models
  with interactions between AI-writing groups and year indicators to regress
  paper-based interdisciplinarity on collaborator-based interdisciplinarity,
  controlling for author fixed effects and annual collaborator counts. Across
  years, paper-based and collaborator-based interdisciplinarity remain positively
  associated. However, this association is consistently weaker for
  high-AI-writing authors and weakens further after 2022. This pattern suggests
  that the interdisciplinarity of high-AI-writing scientists is less tightly
  linked to the disciplinary diversity of their collaborators. In other words,
  while interdisciplinary collaboration continues to contribute to
  interdisciplinary research, high-AI-writing scientists appear increasingly able
  to integrate knowledge across fields without relying as heavily on
  collaborators from different disciplinary backgrounds. This raises the
  possibility that the observed growth in interdisciplinarity reflects not only
  broader collaboration networks but also a shift at the individual level
  (\cref{fig:collaboration-fields}h).

  Lastly, we examine whether changes in collaboration differ between specialists
  and generalists. Generalists' expertise is distributed relatively evenly across
  multiple fields, while specialists often concentrate within a narrow set of
  fields \cite{Lin2026Longevity}. We operationalize this definition dynamically:
  For each year, we classify authors as specialists if their field concentration
  over the preceding three years is high, defined as a field HHI of 0.5 or above,
  and as generalists otherwise. Under this definition, we observe that more
  authors have steadily become generalists over time (Supplementary
  \cref{si-fig:specialist-generalist}a). Both types of authors show declining
  shares of specialist collaborators and rising shares of generalist
  collaborators (Supplementary \cref{si-fig:specialist-generalist}b-c). However,
  the post-2022 shift is not uniform across scientist groups. Notably, post-2022,
  generalist authors increasingly collaborate with other generalists and less
  often collaborate with specialists, while specialist authors' collaboration
  only followed the pre-2023 trajectories. This pattern suggests that the recent
  expansion of interdisciplinary collaboration also reflects a strengthening
  collaboration structure among scientists who already maintain broader field
  portfolios.

  \subsection{Team roles became more differentiated}

  After examining what scientists study and whom they collaborate with, we next
  ask how scientists divide research tasks within teams before and after 2023. To
  examine this question, we use CRediT author contribution statements, which
  provide a standardized vocabulary for describing author roles: conceptual roles
  (conceptualization, methodology, investigation), technical roles (software,
  formal analysis, visualization, data curation, validation), writing roles
  (writing-original draft, writing-review and editing), and management roles
  (funding acquisition, resources, supervision, project administration)
  \citep{Zhang2026How}. The dataset for this analysis includes 137,120
  multi-author papers from the PLOS journal family and journals indexed in PubMed
  Central that have adopted CRediT roles in contribution statements since 2018.

\begin{figure}[htbp]
    \centering
    \includegraphics[width=1\textwidth]{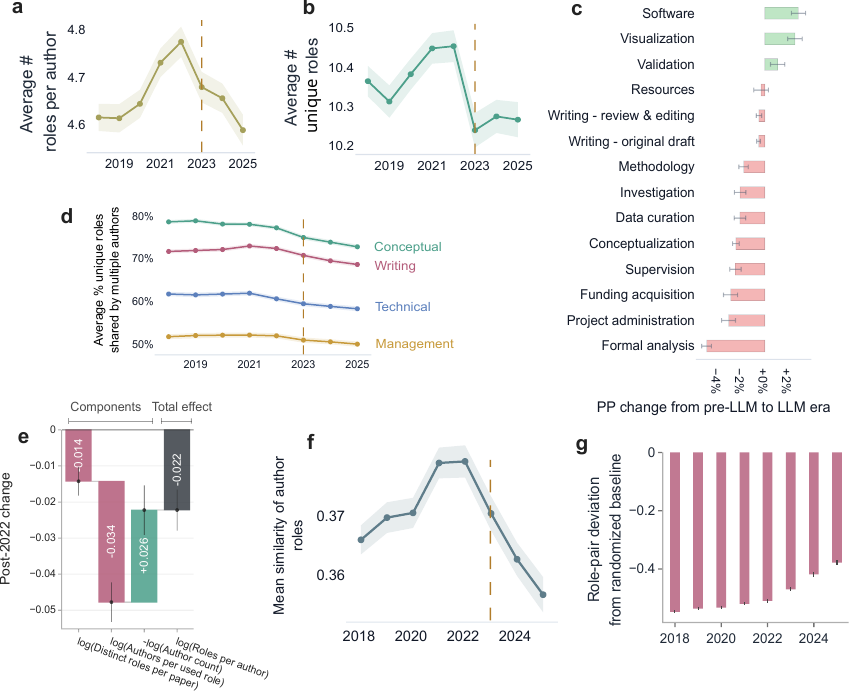}
    \caption{\textbf{Contribution role redistribution and modularity.} Dashed vertical lines mark 2023. Error bands denote 95\% confidence intervals. (a) Average number of roles per author. (b) Average number of unique roles per paper. (c) Percentage-point change in paper-level role presence from the pre-LLM (2020--2022) to LLM-era period (2023--2025) by CRediT role. (d) Average percentage of roles shared by multiple authors, grouped by conceptual, writing, technical, and management role categories. (e) Decomposition of the post-2022 change in average roles per author into components. Effects are changes in log-scale role set measures post-2022 estimated by fixed-effect regression models. (f) Mean similarity of author role sets within papers. (g) Relative percentage deviation from the shuffled random baseline in the role-pair analysis by year.}
    \label{fig:credit-roles}
  \end{figure}

  We first find that individual contributors reported slightly narrower role sets
  after 2022. Among CRediT-reporting papers, the average number of roles per
  author declined from 4.78 in 2022 to 4.56 in 2025 (\cref{fig:credit-roles}a).
  This decline partly reflects a modest reduction in the number of distinct roles
  reported within each paper, which fell from 10.45 in 2022 to 10.27 in 2025
  (\cref{fig:credit-roles}b). However, the pattern is not simply a uniform
  contraction of all contribution categories. From 2020--2022 to 2023--2025, the
  share of papers reporting software increased from 40.6\% to 43.4\%, and the
  share reporting validation increased from 54.1\% to 55.2\%. By contrast,
  several common conceptual, technical, and management roles declined, including
  formal analysis, project administration, funding acquisition, supervision,
  conceptualization, data curation, and investigation (\cref{fig:credit-roles}c).
  A paper-level logistic regression estimated on 41,185 papers from 2023--2025
  shows a similar pattern: after controlling for author count, year fixed
  effects, and paper field fixed effects, a higher AI-writing fraction is
  positively associated with reporting software roles (odds ratio = 1.47, 95\% CI
    [1.31, 1.66]) and validation roles (odds ratio = 1.37, 95\% CI [1.22, 1.54]).
  These results suggest a redistribution of contribution categories rather than a
  simple expansion or contraction of author roles (Supplementary
  \cref{si-fig:si-role-breadth}d).

  The narrowing of individual role sets is driven more by reduced sharing of the
  same roles among coauthors than by the small decline in the number of distinct
  roles reported per paper. As shown in \cref{fig:credit-roles}d, after 2022,
  papers exhibit lower shares of roles that are shared by multiple authors across
  different role categories. Decomposing the post-2022 decline in the average
  number of roles per author further shows that reduced role sharing contributes
  more to this decline than the reduction in the total number of distinct roles
  (\cref{fig:credit-roles}e). This pattern is also visible when coauthors' role
  sets are compared directly. The average pairwise similarity of coauthors' role
  sets increased from 0.371 in 2020 to 0.379 in 2022, but then declined to 0.357
  in 2025. This indicates that coauthors' reported contributions became more
  differentiated after 2022 (\cref{fig:credit-roles}f). The same role-set
  similarity pattern appears when high- and low-AI-writing-rate papers are
  compared (Supplementary \cref{si-fig:si-role-breadth}e), when papers are
  separated by team size (Supplementary \cref{si-fig:si-role-breadth}f), and when
  same-career-stage and different-career-stage author pairs are compared
  (Supplementary \cref{si-fig:si-role-breadth}g).

  Finally, we examine whether author role combinations became less structured
  relative to a randomized within-paper baseline. We reshuffled role assignments
  within each paper while preserving each paper's team size, role composition,
  and author role counts. This procedure provides a null model for the role pairs
  that would be expected if roles were randomly allocated among authors within
  the same paper. We then compared the role pairs actually held by each author
  with those generated under the reshuffled baseline. The resulting deviation
  measure became less negative over time, moving from -0.51 in 2018 to -0.39 in
  2025, with a faster shift after 2022. This indicates that observed role
  combinations moved closer to the randomized baseline, suggesting a weakening of
  fixed role bundling among coauthors. These changes point to a more modular and
  differentiated structure of team-based scientific work in the period of
  widespread LLM diffusion (\cref{fig:credit-roles}g).

  \section{Discussion}
  \label{sec:discussion}

  This study shows that the period after widespread LLM diffusion is associated
  with changes in three connected dimensions of scientific work: broader field
  portfolios and greater research exploration, more interdisciplinary
  collaboration, and a more differentiated labor division within scientific
  teams. The results move beyond the question of whether LLMs changed scientific
  prose. The evidence is descriptive rather than causal, but it indicates that
  LLM diffusion coincided with a broader reorganization of scientific practice.

  The first major finding is that scientists increasingly moved beyond
  single-field exploitation after 2022. Authors' research portfolios became
  broader and more diverse, with greater movement into new and more cognitively
  distant fields. Previous studies have shown that AI tools can improve
  scientists' productivity, research impact, and career development
  \citep{haoArtificialIntelligenceTools2026}; our study extends this literature
  by showing that AI diffusion may also be associated with greater field
  exploration. However, this shift was uneven across the scientific workforce:
  Patterns of LLM usage differ across scientists of career stages. The strongest
  increases in pivot size and Rao--Stirling diversity appear among advanced
  authors. This suggests that while younger scientists tend to have higher
  AI-writing rates (Supplementary \cref{si-fig:si-ai-rate-boxplots}), more
  established scientists may have been especially able to redirect toward new
  areas. Evidence has shown that established tenured professors shift toward
  exploration and novel agendas \cite{Tripodi2025Tenureb}. Their accumulated
  expertise, professional networks, resources, and research autonomy may allow
  them to translate the capabilities of AI tools into more substantial
  intellectual pivots. By contrast, junior scientists may face stronger
  incentives to consolidate a recognizable research agenda and specialize within
  established areas because of funding, publication, and promotion pressures
  \citep{Guo2024Exploration}.

  The pattern is also stronger among non-English low- and middle-income-country
  authors, particularly Chinese authors, indicating that post-2022 tools may have
  reduced some barriers associated with language, search, and access to
  unfamiliar literatures \cite{liuAIAssistedWritingGrowing2025,Liao2024LLMs}.
  While our sample mostly consists of biomedical scientists, scientists from
  engineering and related fields (domain of "Physical Sciences" as in OpenAlex)
  show a more pronounced expansion compared to others, suggesting that the
  "insiders" who are more involved in technical research are more likely to adopt
  the new AI tools and change their research styles \cite{xuAdaptingLLMsHow2025}.

  Moreover, authors' AI-writing rate is positively associated with their
  interdisciplinarity and exploration. Authors with higher AI-writing rates have
  wider field portfolios, more field-spanning collaboration, and larger pivots.
  Comparing the high- and low-AI-writing authors, with basic author features
  matched, high-AI-writing authors are equally productive as low-AI-writing
  authors, but already had broader portfolios and been more exploratory before
  widespread LLM diffusion. This means that selection into AI-assisted writing is
  central to the results; high-AI-writing authors may be more likely to take
  risks and explore new fields and technologies over time. The widening post-2022
  gaps between high- and low-AI-writing authors suggest that LLM-associated
  writing may reinforce existing exploratory tendencies rather than simply create
  them among previously similar scientists. In other words, the results support a
  selection-plus-reinforcement interpretation, not a simple treatment effect.

  The collaboration results show that portfolio expansion was accompanied by
  broader collaboration. After 2022, scientists worked with collaborators from
  more diverse disciplinary backgrounds, and papers more often included
  cross-field collaborators. This finding is consistent with survey findings from
  Elsevier that even in the LLM era, interdisciplinary collaboration remains
  strong and critical to scientists, and scientists still recognize that
  collaboration brings diverse perspectives, fosters creativity, and often is
  necessary to tackle multifaceted research questions that require expertise from
  different areas \citep{ResearcherOfTheFuture}. At the same time, collaboration
  does not fully explain the exploration pattern. The association between
  collaborator-based interdisciplinarity and paper-based interdisciplinarity
  remains positive, but it is weaker among high-AI-writing authors and weakens
  further after 2022. This suggests that some scientists may increasingly
  integrate knowledge across fields without relying as heavily on cross-field
  collaborators. It suggests that boundary-crossing scientific research may be
  increasingly organized through both human collaboration and AI tool-mediated
  individual work \citep{Branda2025Artificial}.

  The specialist-generalist analysis clarifies the social structure of this
  change. The post-2022 shift toward interdisciplinary collaboration is not only
  driven by specialists reaching outward to obtain missing expertise. The
  collaborator pool became more generalist, and generalist authors increasingly
  collaborated with other generalists. It points to strengthening collaboration
  among scientists who already maintain broader field portfolios. This also
  supports the idea that broad-field scientists may become more central in the
  LLM era because they are already positioned to connect multiple literatures and
  tasks.

  Moreover, after 2022, a series of labor division changes occurred within the
  scientific teams. Individuals report fewer contribution roles on average, and
  the unique contribution roles per paper declined modestly. Among all reported
  contribution roles, software and validation tasks increased, while conceptual
  and managing roles decreased. The paper-level AI-writing fraction was also
  associated with the tasks reported in the papers. One possible reason is that
  large language models help automate or accelerate specific research tasks such
  as drafting manuscripts and formulating hypotheses, while scientists are
  responsible for supervising and validating the output
  \cite{zhengAutomationAutonomySurvey2025}.

  In addition, teams appear to report more differentiated task responsibility.
  Team labor became more modular, and coauthors' reported contributions became
  less overlapping and contained fewer fixed pairs. This might indicate that
  individuals are more capable of performing research tasks
  \citep{haoArtificialIntelligenceTools2026}, or that the coordination costs of
  integrating specialized roles and tasks are reduced
  \citep{Burton2024How,Woolley2026Generative}. Together with the aforementioned
  finding that scientists may increasingly integrate knowledge from multiple
  fields without relying as heavily on cross-field collaborators, this pattern
  suggests that human–AI collaboration may partially substitute for some
  functions traditionally provided by collaborators. As AI tools become more
  capable of supporting tasks across specialized domains, scientists may be able
  to overcome functional silos with smaller or differently composed teams,
  thereby reshaping the structure of scientific collaboration
  \citep{dellacquaCyberneticTeammateField2025,Yan2026AIassisted}.

  Overall, the evidence points to a reorganization of scientific work in which
  scientists explore broader field portfolios, collaborate through more diverse
  and more generalist networks, and report a more modular labor division inside
  teams. This reorganization is promising because it suggests that scientists may
  be using LLMs and related AI tools to lower barriers to exploration, coordinate
  across disciplinary boundaries, and redistribute research tasks. However, these
  changes also require caution. Recent work suggests that AI-augmented science
  can expand individual scientists' productivity and visibility while narrowing
  collective scientific focus toward data-rich and already tractable problems
  \citep{haoArtificialIntelligenceTools2026}. Other studies document reliability
  and integrity risks from LLM-generated scholarly content, including fabricated
  or inaccurate references in scientific writing \citep{Chelli2024Hallucination,
    liBibAgentAgenticFramework2026, ZhaoLLM}. Thus, this study should be read as
  evidence of changing scientific organization, not as evidence that LLM
  diffusion necessarily improves research quality. Human verification and
  accountability should remain essential in scientific work.

  \section{Limitations}

  Several limitations shape the interpretation. First, the study is descriptive.
  It does not observe private LLM use and does not estimate a causal effect of
  LLM adoption. The AI-writing measure is a population-level textual proxy. It
  captures language patterns consistent with LLM-assisted writing, but it does
  not identify which author used a tool, which task was assisted, or whether tool
  use caused any portfolio, collaboration, or role change. The exploration and
  collaboration analyses rely on PMC full text linked to OpenAlex, which
  overrepresents biomedical and journal papers. The labor-division analysis
  relies on CRediT-reporting journals, especially PLOS and selected PMC-indexed
  journals. These journals may not represent other journals with distinctive
  contribution-reporting norms. CRediT roles are also reported categories rather
  than direct observations of work. The post-2022 period may reflect several
  changes besides LLM diffusion, including remote collaboration norms after
  COVID-19, funding opportunity changes, and broader transformation of the
  scientific enterprise. These forces may interact with LLM adoption rather than
  operate separately. Due to the selected time window, it remains to be observed
  whether those changes affect the quality and impact of research and
  collaboration. Future research can test whether these patterns generalize
  beyond the current scope.

  \section{Data}

  This study uses two linked datasets. The first dataset is constructed from
  PubMed Central (PMC) full-text records and OpenAlex metadata. PMC provides
  full-text records for indexed publications, which allows us to estimate each
  paper's AI-writing fraction and to build an author-level proxy for exposure to
  LLM-associated writing. We obtained PMC full-text publication data and
  associated metadata in June 2026, matched PMC papers to OpenAlex records
  updated to May 2026 by DOI or PubMed identifier, and retained matched
  publications from 2023 to 2025 as seed papers.

  From these seed papers, we identified 3,959,771 candidate ``seed authors'' and
  retrieved their publication histories from OpenAlex. Retained OpenAlex
  publications were restricted to English-language journal or conference
  publications with a valid DOI. We also excluded papers with more than 20
  authors due to the difficulty of interpreting individual contributions and
  potentially distorted author-level collaboration measures. To support
  author-level temporal comparisons, seed authors were retained only if they
  appeared on at least three matched PMC seed works and had at least three
  OpenAlex publications before 2022. We further excluded the top 2\% of candidate
  seed authors by all-time publication count to reduce the influence of unusually
  prolific authors. The final dataset contains 775,323 seed authors. Across their
  full publication histories, these authors produced 26,057,772 works and
  collaborated with 18,499,462 distinct coauthors, including 5,343,865
  publications and 7,521,167 coauthors during 2023--2025 (Supplementary
  \cref{si-tab:si-publications,si-tab:si-authors-collaborators}).

  We assigned each paper to one of 26 fields and one of 254 subfields based on
  its OpenAlex primary topic. For each year, each author was assigned a primary
  field and subfield from their cumulative publication histories, i.e., all
  observed works published through the given year. Given the PMC-based sample
  construction, the findings should be interpreted primarily as evidence from
  biomedical and biomedical-adjacent research rather than from all scientific
  fields.

  The second dataset is constructed from the PLOS journal family and PMC-indexed
  journals that adopted CRediT author contribution statements. All papers have
  full text available. We began the role-based analysis in 2018, the first year
  with sufficient observed CRediT role coverage in our data. We downloaded PLOS
  journal articles from the official website and identified additional non-PLOS
  PMC-indexed journals that started to report CRediT roles after 2018. Because
  the role analysis focuses on labor division within teams, we excluded 2,683
  solo-authored papers. From the contribution statements, each role description
  is normalized and mapped to the 14 standard CRediT categories. Common wording
  variants are mapped to the same category. The final dataset contains 137,120
  multi-author papers, including 128,613 from PLOS journals and 8,507 from other
  PMC-indexed journals, covering 945,989 paper-author observations (Supplementary
  \cref{si-tab:si-publications}). The final dataset is restricted to journals and
  sources with sufficient observed CRediT reporting from 2018 onward, so changes
  after 2022 are less likely to be driven only by newly entering
  contribution-reporting venues.

  \section{Methods}
  \label{sec:methods}

  \textbf{Field distance and interdisciplinarity measures.} Field distances are estimated from OpenAlex works' references. For each field
  $a$, the workflow constructs a binary vector $v_a$ over citing papers that
  equals one when the citing paper references at least one work in $a$. The
  distance between two fields $a$ and $b$ is computed as the cosine distance between
  $v_a$ and $v_b$:

  \begin{equation}
    d(a,b)=1-\frac{v_a \cdot v_b}{\lVert v_a\rVert \lVert v_b\rVert}.
    \label{eq:field-distance}
  \end{equation}

  Fields that are often cited by the same papers are therefore closer to each
  other, while fields that rarely appear together in reference lists are farther
  apart. The same procedure is used for subfields.

  For each author-window, field portfolio shares are calculated from the author's
  focal papers. If $p_{ict}$ is author $i$'s share of papers in field $c$ during
  year $t$, field concentration is measured by Shannon entropy,

  \begin{equation}
    H_{it}  =  -\sum_{c} p_{ict}\log p_{ict},
  \end{equation}

  A higher entropy value means the author's papers are spread more evenly across
  fields. Rao--Stirling interdisciplinarity adds field distance to the
  field-share distribution. For a set of papers or a collaborator pool, the index
  is

  \begin{equation}
    RS=\sum_c \sum_k p_c p_k d(c,k).
    \label{eq:rao-stirling}
  \end{equation}

  where $p_c$ and $p_k$ are field shares and $d(c, k)$ is the distance between
  the two fields. The index increases when work is distributed across more fields
  and when those fields are more distant from one another.

  \textbf{Research Pivot Measures.} Pivot size measures how far a focal
  paper moves away from an author's recent reference profile. For each
  author-paper pair, the focal vector $v_a$ counts the fields or subfields
  represented in the focal paper's references. The prior vector $v_b$
  counts the same categories in the author's papers from the previous
  three publication years. Same-year papers are not included in the prior
  profile. The pivot measure is cosine distance between the focal and
  prior reference vectors using Eq. \ref{eq:field-distance}. Larger values mean that the paper
  draws on a set of references farther from the author's recent research
  base. Annual pivot measures average this quantity across eligible
  papers by the same author.

  \textbf{High- and Low-AI-Writing Author Matching.} High- and low-AI-writing authors were compared using coarsened exact matching within each publication year. The treatment contrast is defined from the
  author-level post-2022 AI-writing rate: authors with rates of at least 0.15 are
  classified as high-AI-writing authors, while authors with rates of at most 0.05
  are classified as low-AI-writing authors; authors between these thresholds are
  excluded from this comparison. Matching is based on four coarsened variables:

  \begin{itemize}
    \item
          The affiliation-country groups. We cross language classification
          (English/non-English) with income classification (high-income/ low- or
          middle-income) to generate four groups. Countries are classified as
          English-language countries when English is an official or widely
          institutionalized language and is commonly used in higher education,
          government, or scientific communication. Countries are classified as
          high-income under the World Bank income classification in the focal
          year.
    \item
          Career stage is measured in the focal publication year as years since
          the author's first observed publication, grouped into early career
          ($\le$5 years), mid-career (6--15 years), and senior ($\ge$16 years).
    \item
          The primary field is the author's focal primary OpenAlex field among
          all observed publications up to the focal year.
    \item
          Trailing productivity is the author's total publication count over the
          focal year and the previous two years. This productivity measure is
          binned within each publication year into relative productivity groups,
          usually low, medium-low, medium-high, and high.
  \end{itemize}

  A matched stratum is therefore a publication-year-specific combination of
  country context, career stage, primary field, and productivity bin. Strata are
  retained only when they contain both high- and low-AI-writing authors. CEM
  weights are then assigned so that the total matched weight of the high- and
  low-AI-writing groups is balanced within retained strata. Balance diagnostics
  are reported in Supplementary \cref{si-tab:si-cem-balance}; Supplementary
  \cref{si-tab:si-field-ai-groups} reports the field distribution of the two
  groups.

  \textbf{Role-Breadth Regression and Decomposition.} We first estimate a uniform post-2022 regression for three paper-level outcomes: $\log(\text{mean roles per author})$, $\log(\text{distinct roles per paper})$ and $\log(\text{authors per used role})$. The model is

  \begin{equation}
    z_p=\alpha+\beta\,\mathrm{Post2022}_p+\lambda_{p}+\mu_{p}+\epsilon_p.
    \label{eq:post-regression}
  \end{equation}

  where $z_p$ is one of the log outcomes for paper $p$, $\mathrm{Post2022}_p$
  indicates papers published in 2023--2025, $\lambda_{p}$ are paper-field
  controls, and $\mu_{p}$ are journal or source controls. The coefficient $\beta$
  is the post-2022 proportional shift in that outcome, holding field and journal
  fixed. A fourth regression for $\log(\text{author count})$ is used to complete
  the accounting decomposition. The decomposition uses the identity

  \begin{equation}
    \begin{aligned}
      \log(\text{mean roles per author})
       & = \log(\text{distinct roles per paper})    \\
       & \quad + \log(\text{authors per used role}) \\
       & \quad - \log(\text{author count}) .
    \end{aligned}
    \label{eq:decomposition}
  \end{equation}

  The decomposition separates whether the decline in roles per author after 2022
  comes from fewer role categories or less sharing of the roles that remain. In
  the Results, this distinction is used to show that narrower individual role
  sets are driven more by reduced role sharing than by a simple disappearance of
  role categories.

  \textbf{Pairwise Role-Set Similarity.} Pairwise role-set similarity measures how much coauthors overlap in their reported CRediT roles. For each pair of role-coded authors $a$ and $b$ on paper $p$, the Jaccard
  similarity is

  \begin{equation}J_{pab}=\frac{|R_{pa}\cap R_{pb}|}{|R_{pa}\cup R_{pb}|}.\label{eq:jaccard}\end{equation}

  where $R_{pa}$ and $R_{pb}$ are the two authors' role sets. The paper-level
  value is the mean across all author pairs on the paper. Lower values mean
  coauthors report more differentiated role sets. The analysis also calculates
  the same statistic for author pairs in the same career stage and for pairs in
  different career stages, using author-position and career-stage metadata where
  available.

  \textbf{Role-Pair Co-Occurrence Structure.} The role-pair analysis examines whether specific CRediT roles are held together by the same contributor more or less often than expected from the paper's own
  role-assignment structure. A role pair is present when at least one contributor
  on a paper reports both roles. The main comparison uses a within-paper shuffle
  baseline. For each multi-author paper, role assignments are reshuffled while
  preserving team size, each author's number of roles, and each role's number of
  assigned authors. This degree-preserving shuffle keeps the paper's role
  composition fixed and asks whether the observed author role bundles differ from
  what would arise if the same roles were randomly allocated among the same
  authors subject to those constraints. For each author-paper row, the analysis
  compares observed role-pair co-occurrence with the shuffled baseline
  distribution and averages the observed-minus-baseline deviation year by
  year. Values closer to zero mean observed role sets are closer to the
  randomized within-paper baseline.

  \backmatter
  \section*{Declarations}
  \textbf{Acknowledgements} We thank Dr. Cassidy R. Sugimoto and Yiling Lin's constructive suggestions.
  \textbf{Competing interests} The authors declare no competing interests. \\
  \textbf{Data availability} All data used in this study are publicly available. Bibliographic metadata and article records can be obtained from OpenAlex (\url{https://openalex.org}), PubMed Central (\url{https://pmc.ncbi.nlm.nih.gov}), and PLOS journal archives (\url{https://journals.plos.org}). \\
  \textbf{Author contributions} Conceptualization: XZ, XH, CN; Data curation: XZ, XH, JL, CN; Formal analysis: XZ, XH, JL, CN; Investigation: XZ, XH, JL; Methodology: XZ, XH, JL, CN; Project administration: CN; Resources: CN; Software: XZ, XH, JL, CN; Supervision: CN; Validation: XZ, XH, JL, CN; Visualization: XZ, XH, JL, CN; Writing – original draft: XZ, XH, JL; Writing – review \& editing: XZ, XH, JL, CN.\\

  \putbib[references]
\end{bibunit}

\clearpage
\linespread{1}\selectfont
\renewcommand{\thefigure}{S\arabic{figure}}
\renewcommand{\thetable}{S\arabic{table}}
\renewcommand{\theequation}{S\arabic{equation}}
\renewcommand{\thepage}{S\arabic{page}}
\renewcommand{\theHfigure}{SI.\arabic{figure}}
\renewcommand{\theHtable}{SI.\arabic{table}}
\renewcommand{\theHequation}{SI.\arabic{equation}}
\setcounter{figure}{0}
\setcounter{table}{0}
\setcounter{equation}{0}
\setcounter{page}{1}

\makeatletter
\def\@extra@binfo{.SI}
\def\@extra@b@citeb{.SI}
\makeatother
\begin{bibunit}[sciencemag]
  \begin{center}
    \section*{Supplementary Materials for\\ \scititle}
    Xiang~Zheng$^{1}$, Xi~Hong$^{1}$, Jialin~Liu$^{1}$, Chaoqun~Ni$^{1\ast}$\\
    \small$^{1}$Information School, University of Wisconsin--Madison, Madison, WI, USA.\\
    \small$^\ast$Corresponding author. Email: chaoqun.ni@wisc.edu
  \end{center}

  \subsubsection*{This PDF file includes:}
  Supplementary Notes 1--\total{suppnote}\\
  Figures S1--S\total{figure}\\
  Tables S1--S\total{table}

  \newpage

  \suppnote{AI-writing fraction}

  We followed the population-level framework based on word frequency proposed by
  Liang et al. \citep{liangQuantifyingLargeLanguage2025} to quantify the extent
  of LLM usage in academic writing by analyzing word distributions. This approach
  models a focal paper as a mixture of human-written and AI-generated sentence
  distributions and estimates the mixing coefficient, that is, the fraction of
  AI-generated sentences, through maximum likelihood estimation (MLE). We adopted
  the following step-by-step procedure.

  For each word $t$, let $\hat{p}_t$ and $\hat{q}_t$ denote its occurrence
  probabilities in human-written and AI-generated sentences, respectively:

  \begin{equation}\hat{p}_t=\frac{\text{Number of human-written sentences containing }t}{\text{Total number of human-written sentences}}.\label{eq:si-human-word}\end{equation}

  \begin{equation}\hat{q}_t=\frac{\text{Number of AI-generated sentences containing }t}{\text{Total number of AI-generated sentences}}.\label{eq:si-ai-word}\end{equation}

  Because the current study focuses on publications from PubMed Central, we used
  the occurrence probabilities estimated from bioRxiv in previous work
  \citep{KobakDelving}, given the clear disciplinary overlap between bioRxiv and
  the biomedical literature analyzed here. The likelihood of a sentence $x_i$
  under the human-written sentence distribution, $P(x_i)$, and the AI-generated
  sentence distribution, $Q(x_i)$, is defined as follows:

  \begin{equation}P(x_i)=\prod_{t\in x_i}\hat{p}_t\prod_{t\notin x_i}(1-\hat{p}_t).\label{eq:si-px}\end{equation}

  \begin{equation}Q(x_i)=\prod_{t\in x_i}\hat{q}_t\prod_{t\notin x_i}(1-\hat{q}_t).\label{eq:si-qx}\end{equation}

  For each sentence $x_i$, the mixture likelihood is calculated as:

  \begin{equation}L_i(\alpha)=(1-\alpha)P(x_i)+\alpha Q(x_i).\label{eq:si-mixture}\end{equation}

  where $\alpha$ denotes the fraction of AI-generated sentences in the document
  and parameterizes the sentence-level mixture likelihood. For a document
  containing $n$ sentences, the log-likelihood $\mathcal{L}(\alpha)$ is given by:

  \begin{equation}\mathcal{L}(\alpha)=\sum_{i=1}^{n}\log[L_i(\alpha)].\label{eq:si-loglik}\end{equation}

  The AI-generated sentence fraction is then estimated by maximizing the
  log-likelihood under the mixture distribution:

  \begin{equation}\hat{\alpha}=\arg\max_{\alpha\in[0,1]}\mathcal{L}(\alpha).\label{eq:si-alpha}\end{equation}

\begin{figure}[htbp]
    \centering
    \includegraphics[width=0.7\textwidth]{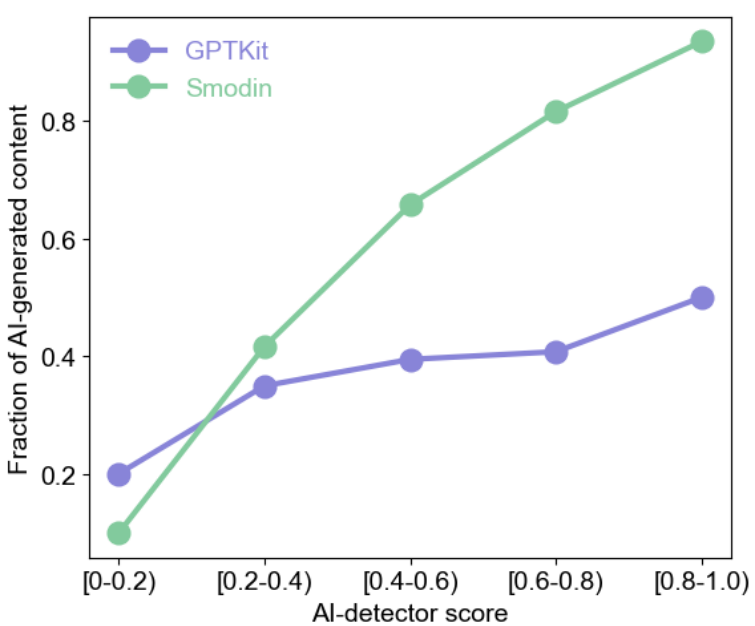}
    \caption{\textbf{Validation of the distribution-based strategy for estimating fraction of AI-generated content.} To validate our estimation strategy, we
      compared its results against two commercial AI detectors, GPTKit and Smodin. We analyzed over 5,000 abstracts from a prior study (14) and grouped them by the AI-generated probability score provided by each detector (x-axis). We then applied our distribution-based strategy to estimate the fraction of AI-generated content for each group    (y-axis). The strong positive correlation shown for both detectors confirms that our strategy's estimates are consistent with established detection tools, supporting its
      effectiveness. Quoted from \citep{liuAIAssistedWritingGrowing2025}.}
    \label{si-fig:ai-detector}
  \end{figure}

  The methodology used in this study is well established and supported by
  multiple layers of validation. This approach has been previously validated and
  applied in prior studies
  \citep{heAcademicJournalsAI2026,KobakDelving,liangQuantifyingLargeLanguage2025,silerDiffusionLargeLanguage2026}.
  Because its statistical estimation is conducted at the corpus level, it is more
  robust than individual-instance inference. Prior test cases showed that the
  model's prediction error for AI-generated content was below 3.5\%, and
  proofreading was found to increase the estimated AI-writing fraction by only
  about 1\% \citep{Liang2024Monitoringa}. We also compared our estimates with
  estimates using online detectors to generate AI probability scores ranging from
  0 to 1 for more than 5,000 abstracts \citep{Liu2024relationship,
    liuAIAssistedWritingGrowing2025}. We grouped these abstracts into probability
  bins, such as {[}0, 0.2), {[}0.2, 0.4), {[}0.4, 0.6), {[}0.6, 0.8), and {[}0.8,
  1.0{]}, and then applied our corpus-level estimation approach to each bin. As
  shown in the validation comparison (Supplementary \cref{si-fig:ai-detector}),
  our estimates are positively correlated with the average AI-generated
  probability based on detection tools, further supporting the effectiveness of
  the strategy used in this study.

  \suppnote{Fixed-effect event-study models}

\begin{figure}[htbp]
    \centering
    \includegraphics[width=1\textwidth]{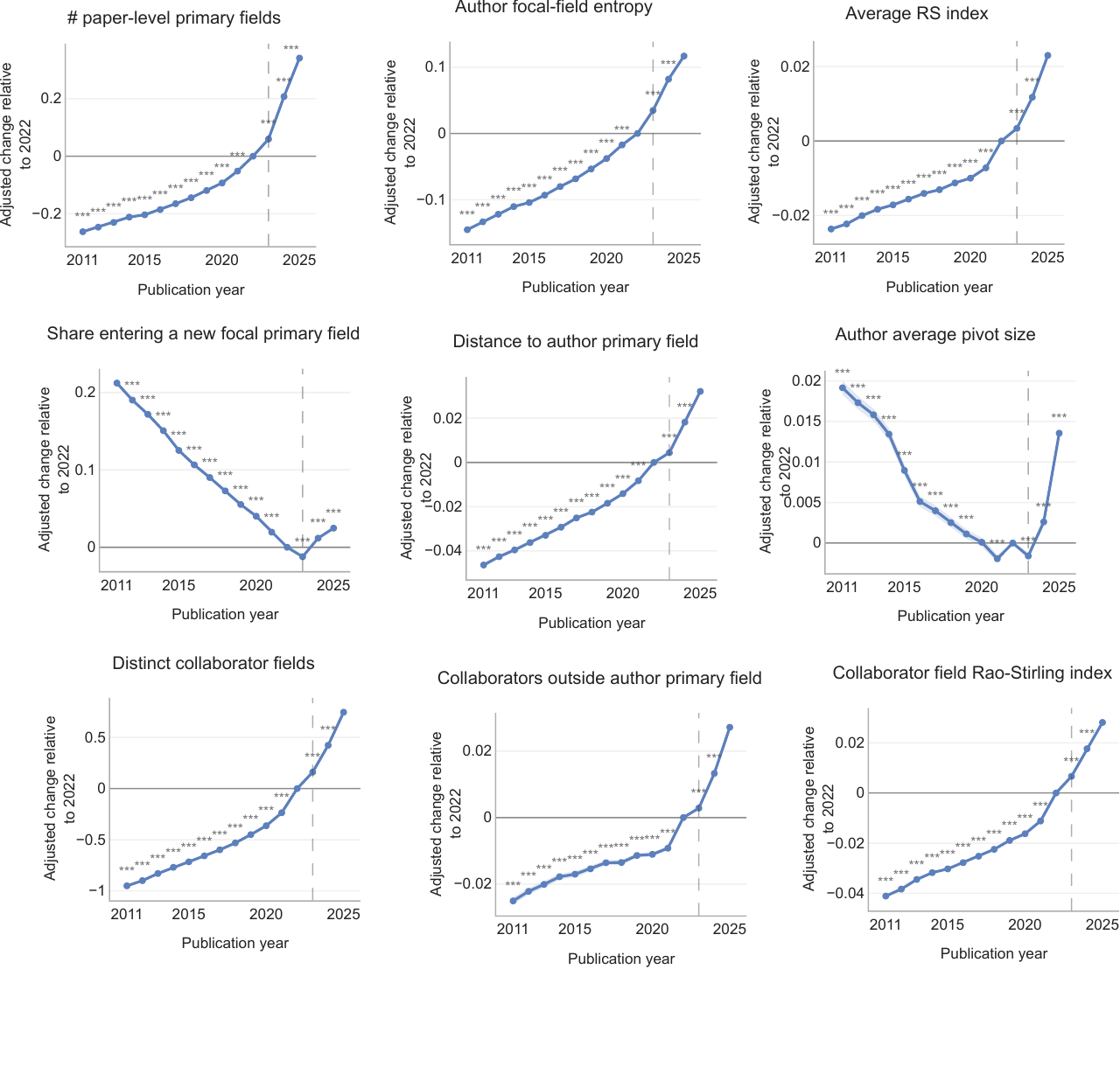}
    \caption{\textbf{Fixed-effect event-study estimates of changes in research exploration and collaboration relative to 2022.} Points show coefficients from author-year regressions that compare each publication year with 2022. Models control for individuals' yearly publication counts and author fixed effects. Standard errors are clustered at author level. Dashed vertical line marks 2023, and asterisks indicate statistical significance. Error bands denote 95\% confidence intervals. The outcome measures are from Figures. 1-2. Positive coefficients indicate that, relative to the same authors' 2022 levels, research portfolios and collaboration networks became broader and more interdisciplinary after widespread LLM diffusion.}
    \label{si-fig:si-event-study}
  \end{figure}

  We additionally examine the exploration, interdisciplinarity, and collaboration
  measures with controls of authors and publication counts. The annual
  event-study models use author-year observations from 2011 through 2025. The
  main within-author specification estimates

  \begin{equation}y_{it}=\alpha_i+n_{it}+\sum_{t\ne 2022}\tau_t\mathbf{1}\{T=t\}+\epsilon_{it}.\label{eq:si-event-study}\end{equation}

  where $y_{it}$ is an author-year outcome, $\alpha_i$ is an author fixed effect
  and 2022 is the reference year, $n_{it}$ is the author's publication count in
  that year. Standard errors are clustered by author. The models confirm that the
  post-2022 increase is not solely driven by differences in the composition of
  authors or publication counts observed across years (Supplementary
  \cref{si-fig:si-event-study}).

\begin{figure}[htbp]
    \centering
    \includegraphics[width=1\textwidth]{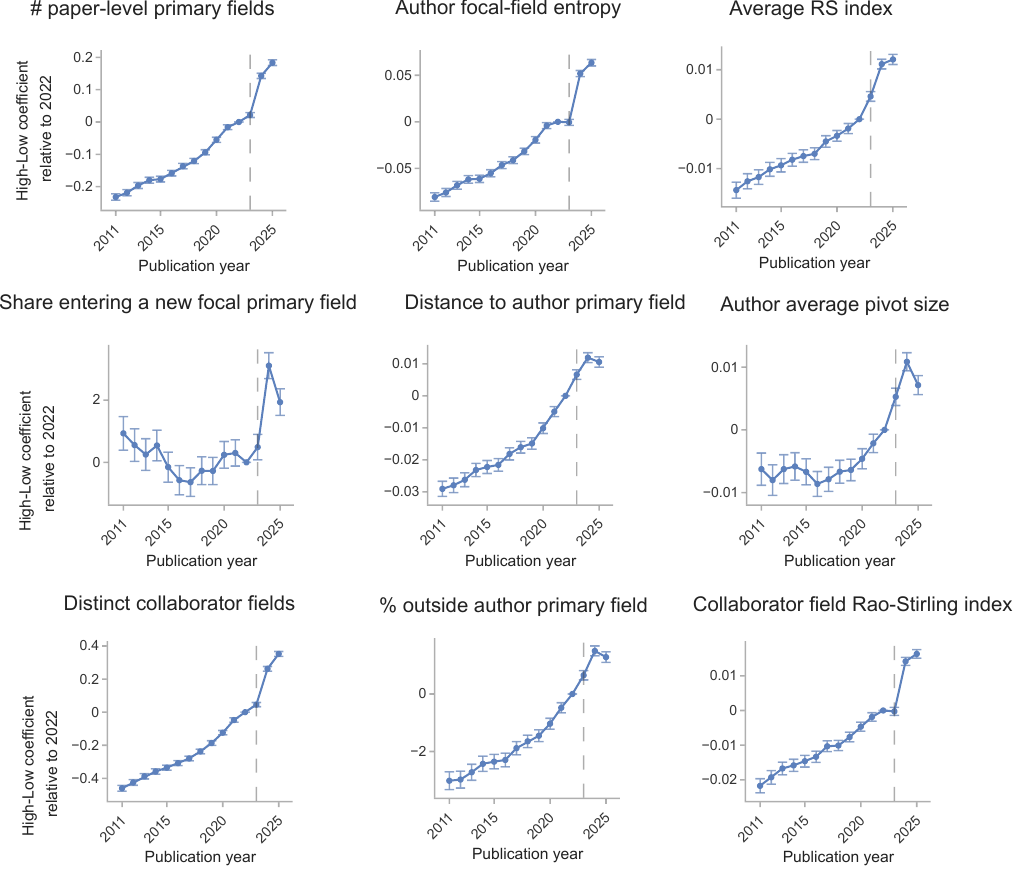}
    \caption{\textbf{High- versus low-AI-writing author event-study estimates.} Points show fixed-effect high-minus-low AI-writing coefficients for each publication year, normalized to 2022; high-AI-writing authors are those with 2023-2025 author AI-writing rates at least 0.15, and low-AI-writing authors are those with rates at most 0.05. Models are based on CEM-matched high/low author-year samples, using CEM weights. Controls include individuals' yearly publication counts and author- and year-level fixed effects. Standard errors are clustered at author level. Dashed vertical line marks 2023. Error bands denote 95\% confidence intervals. The outcome measures are from Figs. 1-2. Positive post-2022 coefficients indicate that high-AI-writing authors increased more than matched low-AI-writing authors, relative to the corresponding high-low difference in 2022.}
    \label{si-fig:si-high-low-event}
  \end{figure}

  The high- versus low-AI-writing event-study models use the same 2022 reference
  year and compare matched high- and low-AI-writing authors. The estimating
  equation includes author fixed effects, year fixed effects, publication counts,
  and interactions between the high-AI-writing indicator and year indicators.
  Regression models are based on CEM-matched high/low author-year samples and
  apply coarsened exact matching (CEM) weights. Balance diagnostics compare the
  pooled author-year covariate distributions before matching with the weighted
  distributions after retaining common-support strata. CEM weighting reduced both
  the L1 imbalance and the maximum absolute category-level standardized mean
  difference to less than 0.001 for each matching variable (Supplementary
  \cref{si-tab:si-cem-balance}). These estimates show whether the high-low
  difference widens or narrows relative to 2022 within the matched sample. The
  results support our findings that the differences between the high- and
  low-AI-writing authors were already evident before 2023 and widened further and
  faster after that year (Supplementary \cref{si-fig:si-high-low-event}).

  \clearpage

  \section*{Supplementary Figures}

  \begin{figure}[p]
    \centering
    \includegraphics[width=1\textwidth]{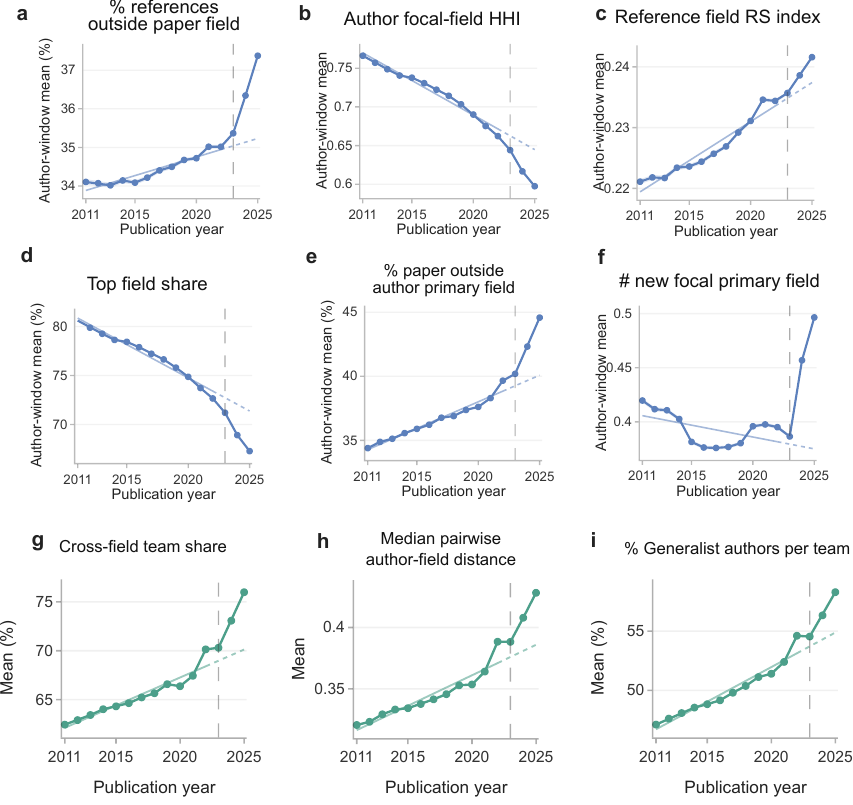}
  \end{figure}
  \figurecaptionpage{\textbf{Alternative measures.} Dashed vertical lines mark 2023. Solid trend lines are linear fits estimated from 2011--2022 values; dashed extensions indicate expected post-2022 values under pre-2023 trends. Error bands denote 95\% confidence intervals. (a) Average share of cited references whose primary field differs from the citing paper's field. This measure is calculated at the paper level and then averaged at the author level. (b) Authors' focal-field Herfindahl-Hirschman index (HHI), which measures how concentrated an author's publications are in a small number of focal fields, with lower values indicating broader field portfolios. (c) Reference-field Rao-Stirling index. This measure captures how diverse and intellectually distant the fields represented in a paper's references are, and then averaged at the author level. (d) Top field share, i.e., the percentage of an author's papers that fall in their most frequent field. (e) Share of an author's papers published outside their established primary field. (f) The number of new primary fields an author enters for the first time. (g) Share of cross-field teams at the paper level that include authors from different primary fields. (h) Median disciplinary distance between pairs of coauthors' primary fields within a team. (i) The average share of generalist team members across papers.}{si-fig:si-alternative-measures}

  \clearpage

  \begin{figure}[htbp]
    \centering
    \includegraphics[width=1\textwidth]{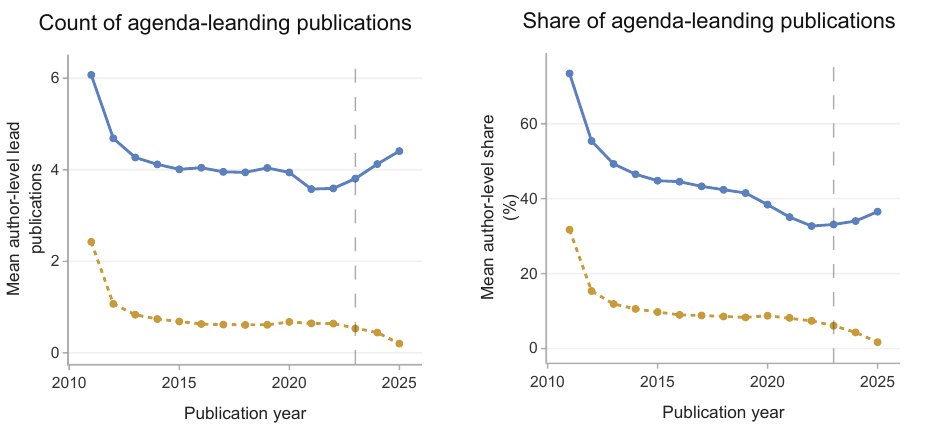}
    \caption{\textbf{Author-level new agenda exploration based on bibliographic coupling.} We create weighted co-reference networks for each scientist based on their publications throughout their career \cite{Zeng2019Increasing}. For each focal author, papers are connected when they cite overlapping references, with edge weights equal to the number of shared references; Louvain communities in this author-specific publication network are treated as research agendas. A publication is coded as a new-agenda lead when it is among the earliest papers in one of the author's detected agendas. The figure reports, by publication year, the mean number (left) and share (right) of new-agenda lead publications per author among authors with at least 10 publications in the full sample. The solid blue line counts isolated papers as one-paper singleton agendas, while the dotted gold line excludes isolated papers before identifying agenda leads. Shaded bands show 95\% confidence intervals across authors, and the dashed vertical line marks 2023. The post-2022 rise in the isolate-inclusive series suggests renewed entry into bibliographically distinct agendas, and this pattern is partly driven by single papers that have no overlapping knowledge bases with previous papers.}
    \label{si-fig:si-new-agenda}
  \end{figure}

  \clearpage

  \begin{figure}[htbp]
    \centering
    \includegraphics[width=1\textwidth]{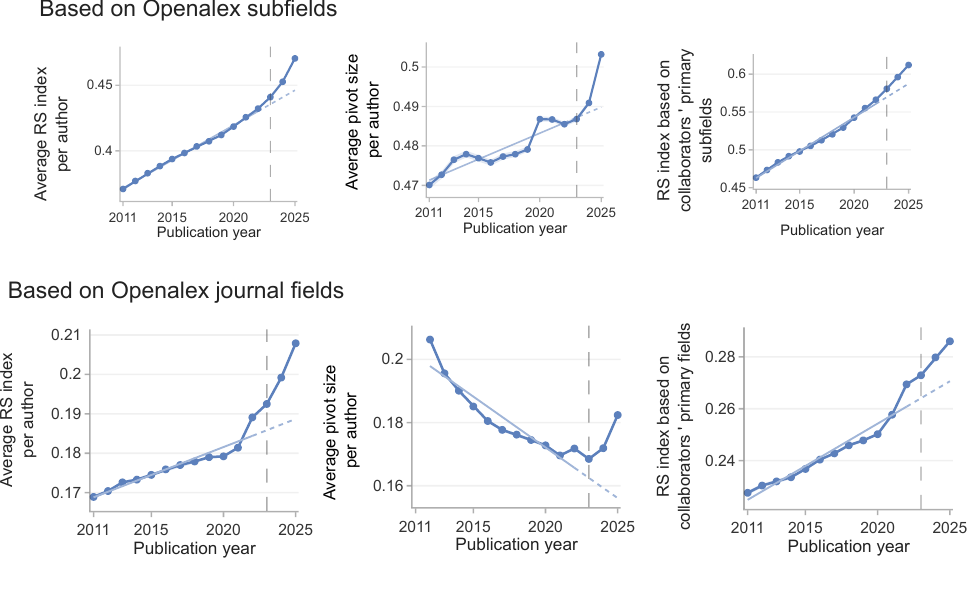}
    \caption{\textbf{Robustness to alternative field definitions.} The figure repeats the measures using subfields in the top row and journal fields in the bottom row. For the journal-based robustness measure, we did not use each paper's own topic assignment. Instead, we assigned fields through the publication source, limited to journals. For each journal, we took the top-ranked OpenAlex topic listed for that source and extracted the corresponding field from the
      topic hierarchy. Each paper and its references were then assigned to the field
      of its publication source. Solid points show annual means, fitted trend lines summarize the pre-2023 trajectory, and dashed vertical lines mark 2023.}
    \label{si-fig:si-field-definitions}
  \end{figure}

  \clearpage

  \begin{figure}[htbp]
    \centering
    \includegraphics[width=1\textwidth]{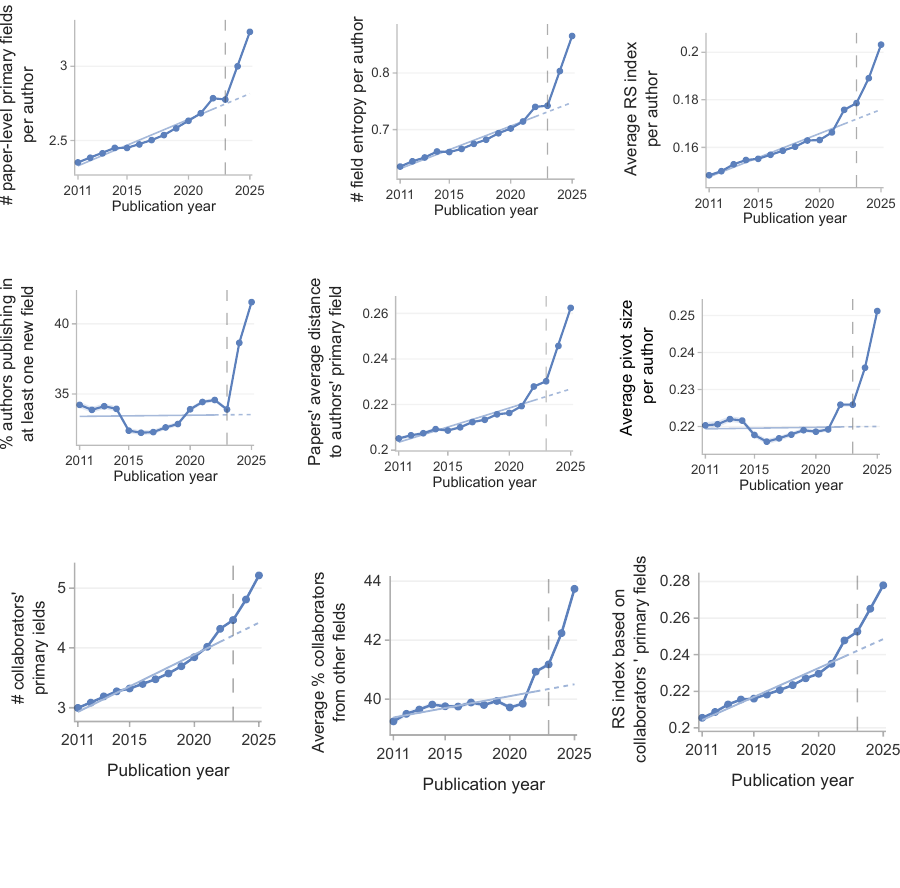}
    \caption{\textbf{Restricted-sample trends.} The sample is restricted to author-years with at least three published papers for paper-based measures and at least three collaborators for collaboration-based measures. The dashed vertical line marks 2023; solid trend lines are fitted from 2011-2022 values, and dashed extensions show the expected post-2022 path under the pre-2023 trend. The outcome measures are from Figs. 1-2.}
    \label{si-fig:si-restricted-sample}
  \end{figure}

  \clearpage

  \clearpage

  \begin{figure}[p]
    \centering
    \includegraphics[width=1\textwidth]{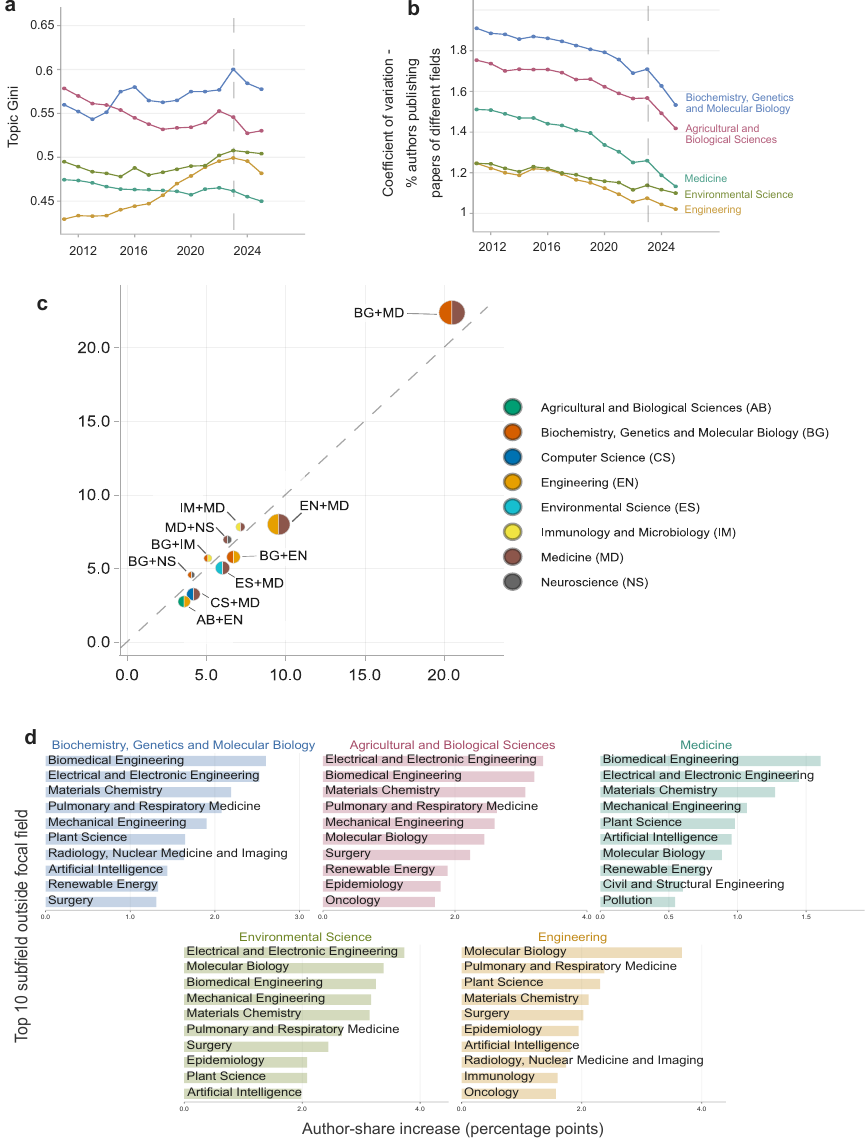}
  \end{figure}
  \figurecaptionpage{\textbf{Topic diversity and field-pair author shares.} The dashed vertical lines in panels a and b mark 2023. (a) Topic Gini index. Each paper is assigned to a primary topic by OpenAlex. This figure contains 3,731 topics that appear as primary topics every year in our sample. It measures how concentrated publications within a field are across primary topics; higher values indicate that papers are more concentrated in fewer topics, while lower values indicate broader topic dispersion. (b) The coefficient of variation measures dispersion in the share of authors publishing in different fields within each broad field; lower values indicate that cross-field publishing is more evenly distributed rather than concentrated among a small subset of authors. (c) Five field-pair combinations with the largest increases and the five with the largest decreases in author share from 2020-2022 to 2023-2025. Each point is an unordered pair of paper primary fields, with the x-axis showing the share of authors publishing in that field pair during 2023-2025 and the y-axis showing the corresponding share during 2020-2022; points below the 45-degree line indicate field pairs that became more common after 2022. Point colors identify the two fields in each pair, and larger points represent larger absolute changes in author share. (d) Top 10 subfields with the largest increases in author shares by field, 2020–2022 to 2023–2025. Each panel ranks the 10 subfields outside the indicated focal field with the largest rise in the share of distinct authors published in that field. Values are percentage-point differences between author shares in 2023–2025 and 2020–2022.}{si-fig:si-topic-diversity}

  \clearpage

  \begin{figure}[htbp]
    \centering
    \includegraphics[width=1\textwidth]{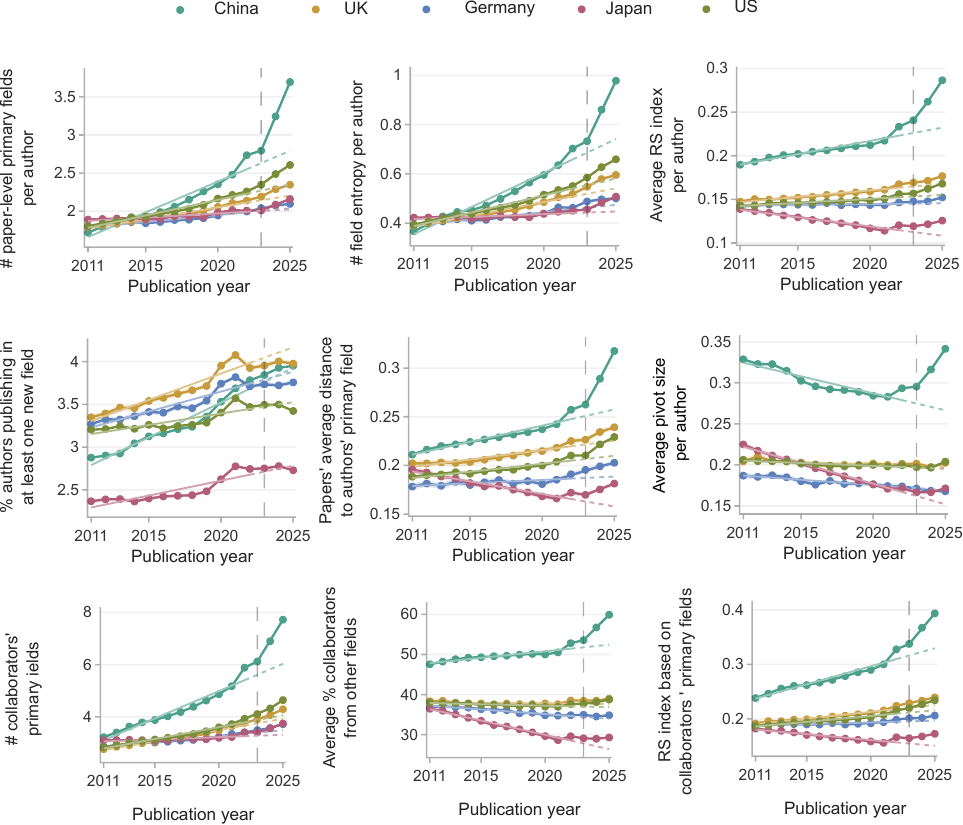}
    \caption{\textbf{Trends by authors' affiliation country.} Five representative countries are shown: China, US, Japan, Germany, and UK. The dashed vertical line marks 2023; solid trend lines are fitted from 2011-2022 values, and dashed extensions show the expected post-2022 path under the pre-2023 trend. The outcome measures are from Figs. 1-2.}
    \label{si-fig:si-top-country-trends}
  \end{figure}

  \clearpage

  \begin{figure}[htbp]
    \centering
    \includegraphics[width=1\textwidth]{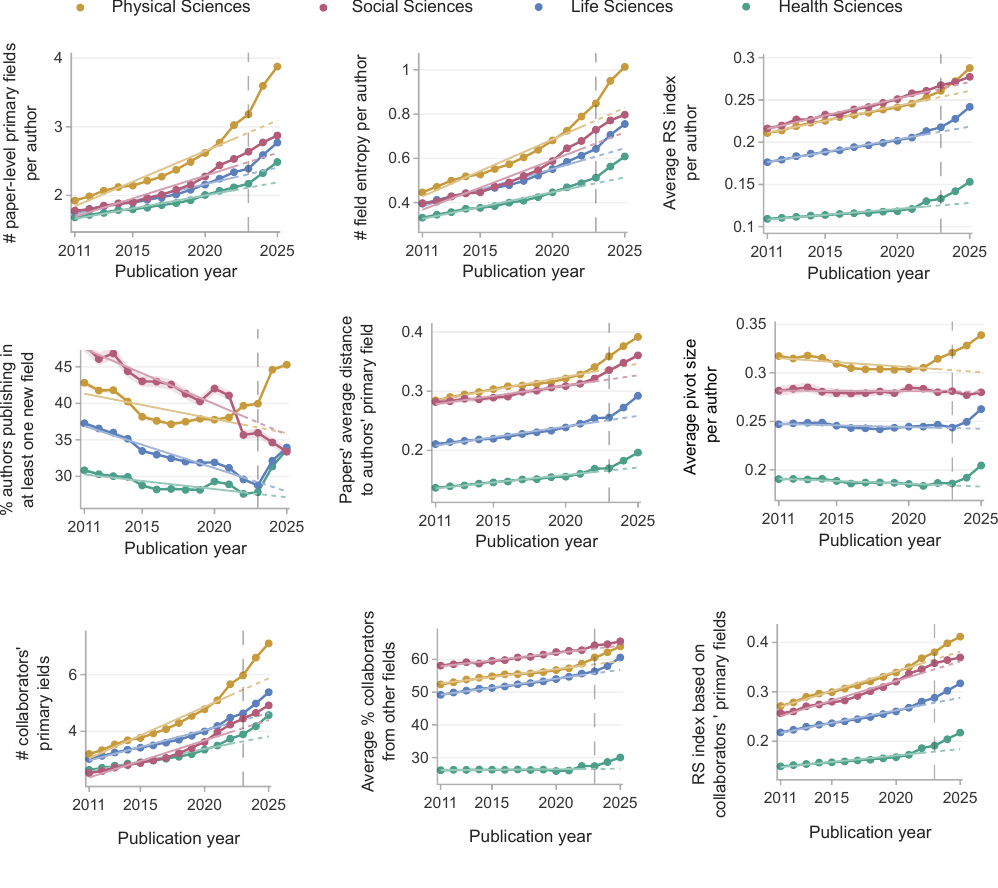}
    \caption{\textbf{Trends by authors' primary domain.} An author's primary domain in a given year is assigned by the most frequent domain in that year's publications. The dashed vertical line marks 2023; solid trend lines are fitted from 2011-2022 values, and dashed extensions show the expected post-2022 path under the pre-2023 trend. The outcome measures are from Figs. 1-2. See the OpenAlex Fields documentation (https://developers.openalex.org/api-reference/fields) for the domain-field mapping.}
    \label{si-fig:si-domain-trends}
  \end{figure}

  \clearpage

  \clearpage

  \begin{figure}[htbp]
    \centering
    \includegraphics[width=1\textwidth]{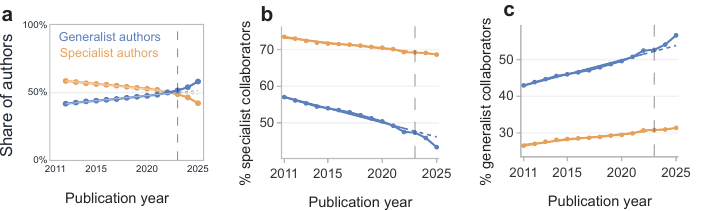}
    \caption{\textbf{Specialist and generalist collaboration.} The dashed vertical lines in panels a and b mark 2023. (a) Share of authors classified as generalists or specialists over time. (b) Percentage of collaborators classified as specialists over time, categorized by the focal authors’ classification. (c) Percentage of collaborators classified as generalists over time, categorized by the focal authors’ classification. Dashed vertical lines mark the post-2022 period.}
    \label{si-fig:specialist-generalist}
  \end{figure}

  \clearpage
  \begin{figure}[htbp]
    \centering
    \includegraphics[width=1\textwidth]{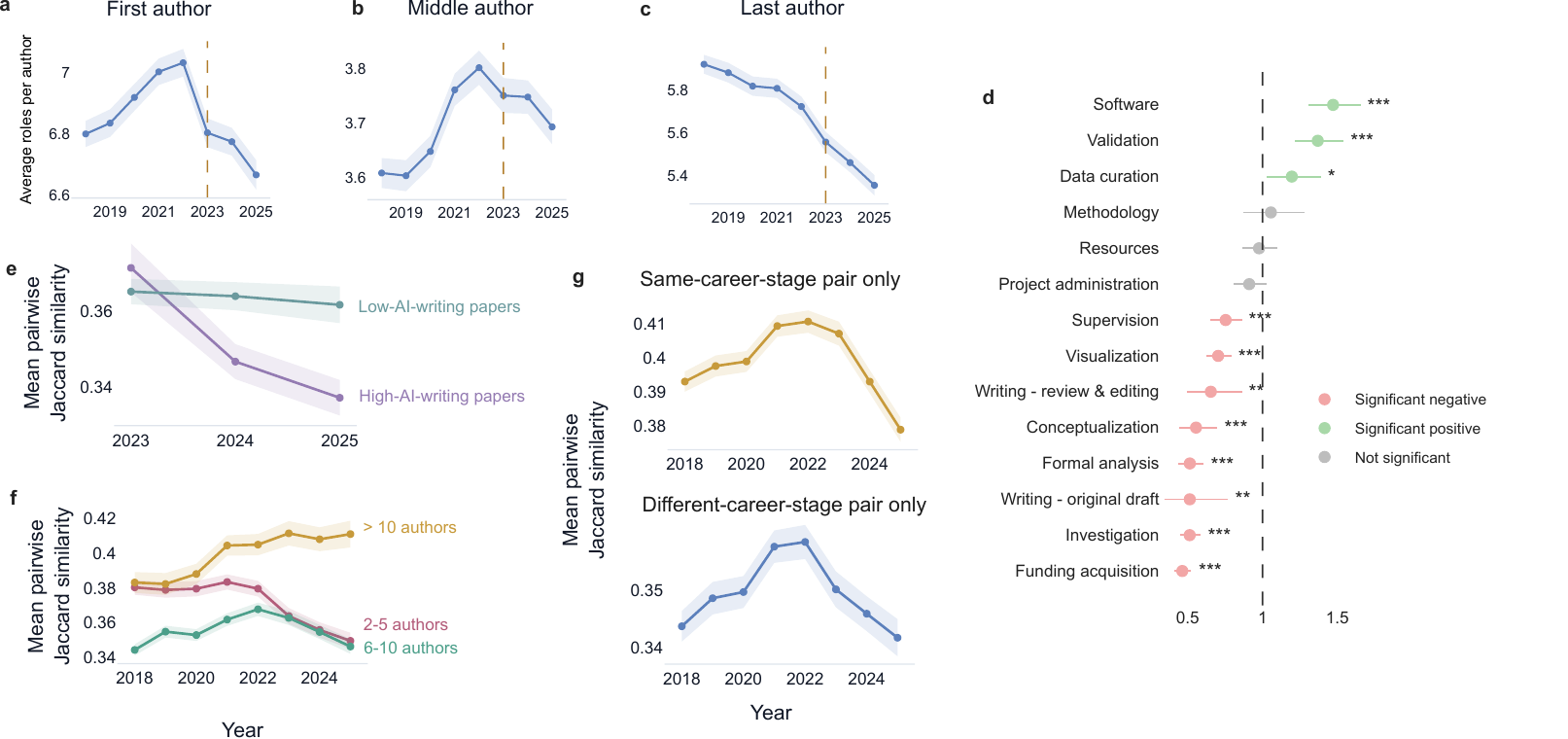}
    \caption{\textbf{CRediT role breadth, role sharing, and AI-writing rate.} (a-c) The average number of reported CRediT roles per author by byline position for first, middle, and last authors; lower values after 2023 indicate narrower individual role sets. (d) Odds ratios from role-specific logistic regressions predicting whether each role is present on a paper, with paper AI-writing rate as the focal predictor and controls for author count, publication year, paper primary field, and publication venue. Values above one indicate a positive association. (e-g) Mean pairwise Jaccard similarity between coauthors' role sets, where higher values mean more role sharing and lower values mean a more differentiated division of labor. (e) compares high-AI-writing-rate papers ($\geq$0.15) with low-AI-writing-rate papers ($\leq$0.05), (f) compares role-set similarity by team size, and (g) separates same-career-stage from different-career-stage author pairs. Shaded bands and horizontal intervals denote 95\% confidence intervals; the dashed vertical line marks 2023.}
    \label{si-fig:si-role-breadth}
  \end{figure}

  \clearpage

  \begin{figure}[htbp]
    \centering
    \includegraphics[width=1\textwidth,height=0.72\textheight,keepaspectratio]{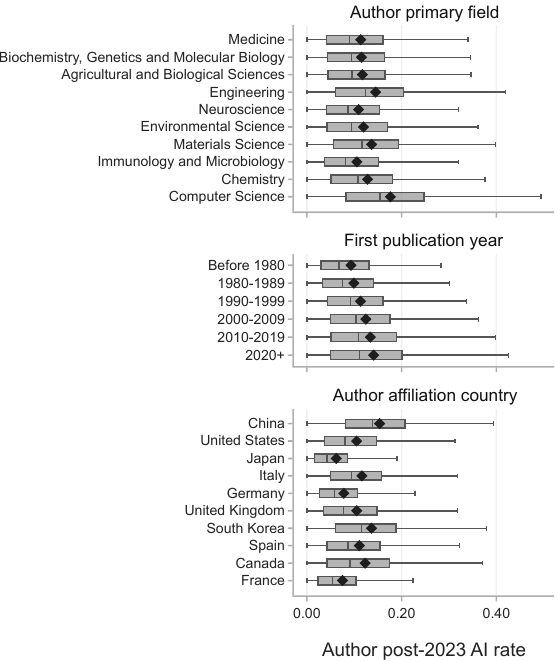}
    \caption{\textbf{Distribution of seed-author AI-writing rates by primary field, first publication year, and affiliation country.} Box plots summarize author-level AI-writing rates, defined as each seed author's mean AI-writing fraction across scored publications in 2023--2025. Panels show the top 10 primary fields by seed-author count, binned first observed publication year, and the top 10 affiliation countries by seed-author count. Diamonds mark group means.}
    \label{si-fig:si-ai-rate-boxplots}
  \end{figure}

  \clearpage

  \clearpage

  \section*{Supplementary Tables}

  \noindent\textbf{Supplementary Table~S\number\numexpr\value{table}+1\relax. Distribution of analytical publications.}

  \begin{longtable}[]{@{}
    >{\raggedright\arraybackslash}p{(\linewidth - 12\tabcolsep) * \real{0.1810}}
    >{\raggedright\arraybackslash}p{(\linewidth - 12\tabcolsep) * \real{0.1469}}
    >{\raggedright\arraybackslash}p{(\linewidth - 12\tabcolsep) * \real{0.1134}}
    >{\raggedright\arraybackslash}p{(\linewidth - 12\tabcolsep) * \real{0.1353}}
    >{\raggedright\arraybackslash}p{(\linewidth - 12\tabcolsep) * \real{0.1251}}
    >{\raggedright\arraybackslash}p{(\linewidth - 12\tabcolsep) * \real{0.1190}}
    >{\raggedright\arraybackslash}p{(\linewidth - 12\tabcolsep) * \real{0.1251}}@{}}
    \toprule\noalign{}
    \begin{minipage}[t]{\linewidth}\raggedright
    \end{minipage}  & \begin{minipage}[t]{\linewidth}\raggedright
                        \textbf{Distinct lifetime papers}
                      \end{minipage} & \begin{minipage}[t]{\linewidth}\raggedright
                                         \textbf{\%}
                                       \end{minipage} & \begin{minipage}[t]{\linewidth}\raggedright
                                                          \textbf{PMC-indexed seed papers}
                                                        \end{minipage} & \begin{minipage}[t]{\linewidth}\raggedright
                                                                           \textbf{\%}
                                                                         \end{minipage} & \begin{minipage}[t]{\linewidth}\raggedright
                                                                                            \textbf{Role-analysis papers}
                                                                                          \end{minipage} & \begin{minipage}[t]{\linewidth}\raggedright
                                                                                                             \textbf{\%}
                                                                                                           \end{minipage}                                                                                                                                            \\
    \midrule\noalign{}
    \endhead
    \bottomrule\noalign{}
    \endlastfoot
    \label{si-tab:si-publications}
    \textbf{Publication year}                    & \textbf{~}                                  & \textbf{~}                                  & \textbf{~}                                  &
    \textbf{~}                                   & \textbf{~}                                  & \textbf{~}                                                                                                                                                                                      \\
    Before 2011                                  & 6,768,373                                   & 26.20\%                                     & /                                           & /                                           & /                                           & /       \\
    2011                                         & 743,175                                     & 2.90\%                                      & /                                           & /                                           & /                                           & /       \\
    2012                                         & 812,324                                     & 3.10\%                                      & /                                           & /                                           & /                                           & /       \\
    2013                                         & 892,627                                     & 3.50\%                                      & /                                           & /                                           & /                                           & /       \\
    2014                                         & 958,062                                     & 3.70\%                                      & /                                           & /                                           & /                                           & /       \\
    2015                                         & 990,550                                     & 3.80\%                                      & /                                           & /                                           & /                                           & /       \\
    2016                                         & 1,056,718                                   & 4.10\%                                      & /                                           & /                                           & /                                           & /       \\
    2017                                         & 1,123,234                                   & 4.40\%                                      & /                                           & /                                           & /                                           & /       \\
    2018                                         & 1,209,534                                   & 4.70\%                                      & /                                           & /                                           & 18,674                                      & 13.60\% \\
    2019                                         & 1,304,414                                   & 5.10\%                                      & /                                           & /                                           & 16,524                                      & 12.10\% \\
    2020                                         & 1,447,056                                   & 5.60\%                                      & /                                           & /                                           & 17,548                                      & 12.80\% \\
    2021                                         & 1,539,842                                   & 6.00\%                                      & /                                           & /                                           & 17,854                                      & 13.00\% \\
    2022                                         & 1,610,685                                   & 6.20\%                                      & /                                           & /                                           & 16,816                                      & 12.30\% \\
    2023                                         & 1,640,013                                   & 6.40\%                                      & 382,767                                     & 35.00\%                                     & 15,838                                      & 11.60\% \\
    2024                                         & 1,803,493                                   & 7.00\%                                      & 408,703                                     & 37.40\%                                     & 17,773                                      & 13.00\% \\
    2025                                         & 1,900,359                                   & 7.40\%                                      & 301,188                                     & 27.60\%                                     & 16,093                                      & 11.70\% \\
    \textbf{Primary field}                       & \textbf{~}                                  & \textbf{~}                                  & \textbf{~}                                  &
    \textbf{~}                                   & \textbf{~}                                  & \textbf{~}                                                                                                                                                                                      \\
    Medicine                                     & 9,174,632                                   & 35.60\%                                     & 492,305                                     & 45.10\%                                     & 52,627                                      & 38.40\% \\
    Biochemistry, Genetics and Molecular Biology & 3,206,494                                   & 12.40\%                                     &
    156,525                                      & 14.30\%                                     & 19,493                                      & 14.20\%                                                                                                                                           \\
    Engineering                                  & 2,681,236                                   & 10.40\%                                     & 54,969                                      & 5.00\%                                      & 4,484                                       & 3.30\%  \\
    Agricultural and Biological Sciences         & 1,363,594                                   & 5.30\%                                      & 64,214                                      &
    5.90\%                                       & 8,453                                       & 6.20\%                                                                                                                                                                                          \\
    Materials Science                            & 1,295,025                                   & 5.00\%                                      & 23,654                                      & 2.20\%                                      & 586                                         &
    0.40\%                                                                                                                                                                                                                                                                                       \\
    Environmental Science                        & 1,167,113                                   & 4.50\%                                      & 38,189                                      & 3.50\%                                      & 7,613                                       &
    5.60\%                                                                                                                                                                                                                                                                                       \\
    Neuroscience                                 & 859,874                                     & 3.30\%                                      & 40,549                                      & 3.70\%                                      & 5,896                                       & 4.30\%  \\
    Chemistry                                    & 740,420                                     & 2.90\%                                      & 10,504                                      & 1.00\%                                      & 339                                         & 0.20\%  \\
    Computer Science                             & 629,390                                     & 2.40\%                                      & 17,880                                      & 1.60\%                                      & 3,338                                       &
    2.40\%                                                                                                                                                                                                                                                                                       \\
    Immunology and Microbiology                  & 625,127                                     & 2.40\%                                      & 31,538                                      & 2.90\%                                      & 5,445
                                                 & 4.00\%                                                                                                                                                                                                                                        \\
    Other                                        & 4,057,554                                   & 15.70\%                                     & 162,331                                     & 14.90\%                                     & 28,846                                      & 21.00\% \\
    \textbf{Source type}                         & \textbf{~}                                  & \textbf{~}                                  & \textbf{~}                                  & \textbf{~}
                                                 & \textbf{~}                                  & \textbf{~}                                                                                                                                                                                      \\
    Journal                                      & 25,571,690                                  & 99.10\%                                     & 1,092,656                                   & 100.00\%                                    & 137,120                                     &
    100.00\%                                                                                                                                                                                                                                                                                     \\
    Conference                                   & 228,769                                     & 0.90\%                                      & 2                                           & \textless0.1\%                              & 0                                           & 0.00\%  \\
  \end{longtable}

  \clearpage

  \noindent\textbf{Supplementary Table~S\number\numexpr\value{table}+1\relax. Distribution of authors and their collaborators. Author attributes are from their latest occurrence in the analytical data.}

  \begin{longtable}[]{@{}
    >{\raggedright\arraybackslash}p{(\linewidth - 8\tabcolsep) * \real{0.4079}}
    >{\raggedright\arraybackslash}p{(\linewidth - 8\tabcolsep) * \real{0.1670}}
    >{\raggedright\arraybackslash}p{(\linewidth - 8\tabcolsep) * \real{0.1039}}
    >{\raggedright\arraybackslash}p{(\linewidth - 8\tabcolsep) * \real{0.2174}}
    >{\raggedright\arraybackslash}p{(\linewidth - 8\tabcolsep) * \real{0.1039}}@{}}
    \toprule\noalign{}
    \begin{minipage}[b]{\linewidth}\raggedright
    \end{minipage}  & \begin{minipage}[b]{\linewidth}\raggedright
                        \textbf{Distinct authors}
                      \end{minipage} & \begin{minipage}[b]{\linewidth}\raggedright
                                         \textbf{\%}
                                       \end{minipage} & \begin{minipage}[b]{\linewidth}\raggedright
                                                          \textbf{Distinct collaborators}
                                                        \end{minipage} & \begin{minipage}[b]{\linewidth}\raggedright
                                                                           \textbf{\%}
                                                                         \end{minipage}                                                                                  \\
    \midrule\noalign{}
    \endhead
    \bottomrule\noalign{}
    \endlastfoot
    \label{si-tab:si-authors-collaborators}
    Primary field                                &                                             &                                             &                                             &         \\
    Medicine                                     & 432,045                                     & 55.7\%                                      & 7,131,425                                   & 38.5\%  \\
    Biochemistry, Genetics and Molecular Biology & 108,028                                     & 13.9\%                                      &
    2,193,322                                    & 11.9\%                                                                                                                                            \\
    Engineering                                  & 47,380                                      & 6.1\%                                       & 2,059,277                                   & 11.1\%  \\
    Agricultural and Biological Sciences         & 47,235                                      & 6.1\%                                       & 1,223,510                                   &
    6.6\%                                                                                                                                                                                            \\
    Environmental Science                        & 22,663                                      & 2.9\%                                       & 807,270                                     & 4.4\%   \\
    Materials Science                            & 18,433                                      & 2.4\%                                       & 666,025                                     & 3.6\%   \\
    Neuroscience                                 & 18,162                                      & 2.3\%                                       & 396,217                                     & 2.1\%   \\
    Psychology                                   & 13,963                                      & 1.8\%                                       & 378,162                                     & 2.0\%   \\
    Immunology and Microbiology                  & 11,453                                      & 1.5\%                                       & 383,237                                     & 2.1\%   \\
    Computer Science                             & 9,617                                       & 1.2\%                                       & 546,123                                     & 3.0\%   \\
    Other                                        & 46,344                                      & 6.0\%                                       & 2,714,894                                   & 14.7\%  \\
    First publication year                       &                                             &                                             &                                             &         \\
    Before 1980                                  & 50,616                                      & 6.50\%                                      & 948,214                                     & 5.10\%  \\
    1980-1989                                    & 56,742                                      & 7.30\%                                      & 861,384                                     & 4.70\%  \\
    1990-1999                                    & 136,430                                     & 17.60\%                                     & 1,689,411                                   & 9.10\%  \\
    2000-2009                                    & 238,162                                     & 30.70\%                                     & 3,137,814                                   & 17.00\% \\
    2010-2019                                    & 257,163                                     & 33.20\%                                     & 5,947,295                                   & 32.10\% \\
    2020+                                        & 36,210                                      & 4.70\%                                      & 5,915,325                                   & 32.00\% \\
    Affiliation country                          &                                             &                                             &                                             &         \\
    China                                        & 211,262                                     & 27.2\%                                      & 2,942,484                                   & 15.9\%  \\
    United States                                & 114,418                                     & 14.8\%                                      & 2,847,651                                   & 15.4\%  \\
    Japan                                        & 35,640                                      & 4.6\%                                       & 746,118                                     & 4.0\%   \\
    Italy                                        & 31,360                                      & 4.0\%                                       & 410,825                                     & 2.2\%   \\
    Germany                                      & 28,773                                      & 3.7\%                                       & 593,239                                     & 3.2\%   \\
    United Kingdom                               & 26,770                                      & 3.5\%                                       & 586,737                                     & 3.2\%   \\
    South Korea                                  & 20,784                                      & 2.7\%                                       & 414,113                                     & 2.2\%   \\
    Spain                                        & 17,063                                      & 2.2\%                                       & 336,247                                     & 1.8\%   \\
    Canada                                       & 15,357                                      & 2.0\%                                       & 349,663                                     & 1.9\%   \\
    France                                       & 15,182                                      & 2.0\%                                       & 388,736                                     & 2.1\%   \\
    Other                                        & 258,714                                     & 33.4\%                                      & 8,883,649                                   & 48.0\%  \\
  \end{longtable}

  \clearpage

  \noindent\textbf{Supplementary Table~S\number\numexpr\value{table}+1\relax. Distribution of primary fields by author AI-writing group. Shares are within each author AI-writing group; the middle AI-writing-rate band is excluded.}

  \begin{longtable}[]{@{}
    >{\raggedright\arraybackslash}p{(\linewidth - 8\tabcolsep) * \real{0.3636}}
    >{\raggedright\arraybackslash}p{(\linewidth - 8\tabcolsep) * \real{0.1705}}
    >{\raggedright\arraybackslash}p{(\linewidth - 8\tabcolsep) * \real{0.1477}}
    >{\raggedright\arraybackslash}p{(\linewidth - 8\tabcolsep) * \real{0.1705}}
    >{\raggedright\arraybackslash}p{(\linewidth - 8\tabcolsep) * \real{0.1477}}@{}}
    \toprule\noalign{}
    \begin{minipage}[b]{\linewidth}\raggedright
      \textbf{Field}
    \end{minipage}  & \begin{minipage}[b]{\linewidth}\raggedright
                        \textbf{High AI authors}
                      \end{minipage} & \begin{minipage}[b]{\linewidth}\raggedright
                                         \textbf{High AI share}
                                       \end{minipage} & \begin{minipage}[b]{\linewidth}\raggedright
                                                          \textbf{Low AI authors}
                                                        \end{minipage} & \begin{minipage}[b]{\linewidth}\raggedright
                                                                           \textbf{Low AI share}
                                                                         \end{minipage}                                                                                  \\
    \midrule\noalign{}
    \endhead
    \bottomrule\noalign{}
    \endlastfoot
    \label{si-tab:si-field-ai-groups}
    Medicine                                     & 125,903                                     & 51.59\%                                     & 124,146                                     & 58.58\% \\
    Biochemistry, Genetics and Molecular Biology & 31,372                                      & 12.85\%                                     & 31,294                                      & 14.77\% \\
    Engineering                                  & 20,945                                      & 8.58\%                                      & 8,453                                       & 3.99\%  \\
    Agricultural and Biological Sciences         & 13,130                                      & 5.38\%                                      & 13,592                                      & 6.41\%  \\
    Environmental Science                        & 6,954                                       & 2.85\%                                      & 6,599                                       & 3.11\%  \\
    Materials Science                            & 6,625                                       & 2.71\%                                      & 4,291                                       & 2.02\%  \\
    Neuroscience                                 & 4,182                                       & 1.71\%                                      & 6,013                                       & 2.84\%  \\
    Psychology                                   & 7,376                                       & 3.02\%                                      & 1,498                                       & 0.71\%  \\
    Immunology and Microbiology                  & 2,206                                       & 0.90\%                                      & 4,583                                       & 2.16\%  \\
    Computer Science                             & 5,721                                       & 2.34\%                                      & 1,063                                       & 0.50\%  \\
    Chemistry                                    & 2,599                                       & 1.06\%                                      & 2,214                                       & 1.04\%  \\
    Health Professions                           & 4,242                                       & 1.74\%                                      & 390                                         & 0.18\%  \\
    Social Sciences                              & 3,513                                       & 1.44\%                                      & 573                                         & 0.27\%  \\
    Dentistry                                    & 2,311                                       & 0.95\%                                      & 1,396                                       & 0.66\%  \\
    Physics and Astronomy                        & 1,506                                       & 0.62\%                                      & 1,895                                       & 0.89\%  \\
    Earth and Planetary Sciences                 & 823                                         & 0.34\%                                      & 1,364                                       & 0.64\%  \\
    Energy                                       & 874                                         & 0.36\%                                      & 389                                         & 0.18\%  \\
    Economics, Econometrics and Finance          & 948                                         & 0.39\%                                      & 178                                         & 0.08\%  \\
    Nursing                                      & 491                                         & 0.20\%                                      & 440                                         & 0.21\%  \\
    Business, Management and Accounting          & 798                                         & 0.33\%                                      & 58                                          & 0.03\%  \\
    Mathematics                                  & 493                                         & 0.20\%                                      & 330                                         & 0.16\%  \\
    Veterinary                                   & 147                                         & 0.06\%                                      & 579                                         & 0.27\%  \\
    Pharmacology, Toxicology and Pharmaceutics   & 336                                         & 0.14\%                                      & 297                                         & 0.14\%  \\
    Decision Sciences                            & 256                                         & 0.10\%                                      & 58                                          & 0.03\%  \\
    Chemical Engineering                         & 160                                         & 0.07\%                                      & 106                                         & 0.05\%  \\
    Arts and Humanities                          & 151                                         & 0.06\%                                      & 112                                         & 0.05\%  \\
  \end{longtable}

  \clearpage

  \noindent\textbf{Supplementary Table~S\number\numexpr\value{table}+1\relax. Covariate balance before and after coarsened exact matching.}

  \begin{longtable}[]{@{}
    >{\raggedright\arraybackslash}p{(\linewidth - 8\tabcolsep) * \real{0.3000}}
    >{\centering\arraybackslash}p{(\linewidth - 8\tabcolsep) * \real{0.1750}}
    >{\centering\arraybackslash}p{(\linewidth - 8\tabcolsep) * \real{0.1750}}
    >{\centering\arraybackslash}p{(\linewidth - 8\tabcolsep) * \real{0.1750}}
    >{\centering\arraybackslash}p{(\linewidth - 8\tabcolsep) * \real{0.1750}}@{}}
    \toprule\noalign{}
    \textbf{Matching variable}                &
    \multicolumn{2}{c}{\textbf{L1 imbalance}} &
    \multicolumn{2}{c}{\textbf{Maximum absolute SMD}}                                                            \\
    \cmidrule(lr){2-3}\cmidrule(l){4-5}
                                              & \textbf{Before CEM}   & \textbf{Weighted CEM} &
    \textbf{Before CEM}                       & \textbf{Weighted CEM}                                            \\
    \midrule\noalign{}
    \endhead
    \bottomrule\noalign{}
    \endlastfoot
    \label{si-tab:si-cem-balance}
    Country context                           & 0.320                 & $<0.001$              & 0.632 & $<0.001$ \\
    Career stage                              & 0.120                 & $<0.001$              & 0.241 & $<0.001$ \\
    Primary field                             & 0.139                 & $<0.001$              & 0.180 & $<0.001$ \\
    Trailing three-year productivity bin      & 0.024                 & $<0.001$              & 0.052 & $<0.001$ \\
  \end{longtable}

  \noindent\textit{Note:} The high-AI-writing group includes authors whose
  2023--2025 author-level AI-writing rate is at least 0.15; the low-AI-writing
  group includes authors whose rate is at most 0.05. Matching was conducted
  separately within each publication year on country context, career stage,
  primary field, and trailing three-year productivity bin. L1 imbalance is half
  the sum of the absolute differences between the high- and low-AI-writing groups'
  category proportions. Maximum absolute SMD is the largest absolute
  standardized mean difference across a variable's indicator categories. The
  post-CEM diagnostics use CEM weights in the retained common-support strata.
  Values below 0.001 reflect numerical precision and confirm the expected balance
  on the variables that define the CEM strata.

  \clearpage

  \renewcommand\refname{References and Notes}
  \putbib[references]
\end{bibunit}
\end{document}